# Theory of Sol-Gel Transition in Thermoreversible Gels with due Regard for Fundamental Role of Mesoscopic Cyclization Effects.
## I. Thermodynamic and Structural Characteristics of Gel Phase.

Igor Erukhimovich*, Michael V. Thamm, Alexander V. Ermoshkin

*Moscow State University, Moscow, 117234, Russia*

## ABSTRACT

The sol-gel transition (SGT), upon which the infinite cluster (IC) of thermoreversibly bonded particles (gel fraction) appears against a background of a set of finite clusters (sol fraction), is first quantitatively considered with due regard for large and complicated (mesoscopic) cycles inevitably present in the IC. To this end we present a new approach based on a concept of the monomer identity breaking and density functional description. We strictly derive, via a proper choice of basic structural units of the gel fraction, all statements usually supposed to be just Flory (Stockmayer) assumptions. A further analysis of the IC structure reveals some new IC structural units (those involved into mesoscopic cycles) overlooked in both Flory and Stockmayer approaches and to be described by a new order parameter characteristic only of the gel phase. As a result, the SGT is found to transform from a geometric phenomenon to a genuine 1st order phase transition always followed by a phase separation into sol and gel phases. The free energy, total conversion, volume fractions of the gel fraction and dangling monomers as well as other structural quantities are calculated as functions of a reduced monomer density and analyzed for all the existing models. The Flory approach is found to be superior to the Stockmayer-Tanaka one and satisfactorily describe some of the dense weak gel properties but fail (even qualitatively) in a quite extended vicinity of the SGT.

# 1. Introduction.

The sol-gel transition (SGT), upon which the infinite cluster (IC) of bonded particles (the gel fraction) appears against a background of a set of finite clusters (the sol fraction), is one of the most fascinating phenomena of polymer physics. Flory[1,2] and Stockmayer[3] first understood its profound physical background. They found a distribution $n(k)$ of the number $n$ of clusters consisting of $k$ monomers as well as conditions when the IC occurs in systems of reacting monomers. But what are the structure of the IC and the state of chemical equilibrium between the gel and sol fractions stays to be a matter of controversy until now. Briefly, both Flory and Stockmayer assume the tree-like structure of both finite and infinite clusters for any finite scale. But on the infinite scale i.e. in the thermodynamic limit $V,N \to \infty$, $\rho = N/V = const$ ($V$, $N$ and $\rho$ being the volume of the system, total number of particles and their density, respectively), their IC pictures differ strongly. Whereas the Flory theory resulted in a gel containing some amount of cyclic (but closed at infinity only) paths, Stockmayer claimed that getting such cycles on the base of consideration involving no explicit cycles proves only an inconsistency of these considerations and suggested that, within the tree approximation, no cycles appear in the IC at all. This controversy has been revived in the last decade in connection with the attempts to understand and describe the SGT in weak (annealed) gels,[4] i.e. the systems capable of forming the branched and cycled structures supposed to be in chemical equilibrium as to making and breaking some saturated bonds.

The state of the real high-molecular polymer gels is rather frozen than annealed (that of a chemical equilibrium). However, there are a lot of low-molecular self-associating solvents whose properties influence the phase behavior of the polymer solutions very much. Water, alcohols and silicate melts are well known examples of weak gels which speaks itself how important are these systems for chemistry, biophysics and earth sciences as well as for diverse applications in technology. On the other hand, the behavior of more realistic frozen polymer networks can be described via a replica generalization[5] of the weak gel description that substantially depends on which model is chosen as a basement for such a generalization. So, understanding of the annealed gels behavior is a primary goal for understanding of both the polymer solutions in the associating solvents and the frozen polymer networks.

In the works by Tanaka[6-11] who adopted the Stockmayer approach, the SGT appears to be the genuine 3rd order phase transition. Other authors[12-21] analyzed thermodynamics of clustering in weak gels within a mean-field approach that implies a simple dependence of the conversion, i.e., the fraction of the chemical groups actually participating in the formation of labile saturated bonds, on the total density of all groups. (This dependence is equivalent to the Flory principle of the equal reaction ability). These authors concluded that the SGT is a purely geometric phenomenon leading to no singularities in the thermodynamic behavior of weak gels. In particular, the Tanaka-Stockmayer approach was criticized recently by Semenov and Rubinstein.[21]

However, *none* of the aforementioned theories takes into account properly the most essential property of the infinite networks (apart from the infiniteness itself): *the dominant role of the cyclization effects for description of the IC structure*. Indeed, the correction due to small loops (see Figures 1b,c) was found[22,23] to be proportional to the parameter $\varepsilon = \left( f \rho a^3 \right)^{-1}$. (Here $\rho$ and $f$ are the number density and functionality of the monomers $A_f$, $a$ is a distance (the bond length) between the bonded monomers.) On the other hand, no IC can be embedded into the real 3D space without large and complicated closed paths of chemical bonds (this evident fact is sometimes referred to as the Malthusian paradox). It follows that smallness of $\varepsilon$ implies only, that it is the amount of *small and simple loops* that could be neglected. On the contrary, the amount of the bonds involved into *large complicated cycles* (see Figure 1d) is shown within a proper perturbation procedure[23-25] to increase when the conversion increases above the SGT. Thus, any adequate description of the IC structure must allow for the cyclization effects *explicitly* (rather than *implicitly* as in Flory case).

Such a description was first proposed by Erukhimovich[25] via simultaneous application of the field-theoretical[26] and density functional[27] methods. Unfortunately, this work seems to be not quite understood yet or even misinterpreted,[28] which is partially explained by its composition and complexity. In the present paper we improve the original approach[25] and solve finally the old controversy by giving a unified density functional description of all existing SGT models. To this purpose the presentation is organized to gradually add complexity and accuracy. In Section 2 we define our model and obtain the contribution to the free energy due to formation between the associating particles without invoking any idea of the IC structure. In Section 3 we elaborate the concept of the *broken monomers identity* (BMI) playing the uppermost role in our approach. The idea of this concept is to partition all the monomers into the classes that topologically differ by their location within the gel or sol fraction and characterize all the monomers belonging to each class by a particular coarse-grained density. Then the thermodynamically equilibrium partition is to be determined by the requirement of the minimal (with respect to all possible partitions) free energy. Using the BMI concept we *derive* (rather than *assume*) *all* the Flory and Stockmayer results concerning the SGT. As a result, we show here that both the Stockmayer and Flory gel pictures correspond to particular states of the partial equilibrium with respect to forming some elements of the IC, the Stockmayer state being turned out to be unstable as to transforming into the Flory one. In Section 4, possessing the central place in the present paper, we give a new treatment of the chemical equilibrium between the sol-fraction and strongly cycled gel thus extending our preceding consideration.[25,29] We show that the Flory state itself is unstable as to transforming into a strongly cycled state, containing mesoscopically large but finite closed trails of the chemical bonds. The symmetry difference between the monomers that belong to these trails and those which do not, causes occurrence of a new order parameter peculiar for the gel phase which results, in turn, into violation of the Flory principle of equal reaction ability. We also calculate here the structural free energy of a mesoscopically cycled (MC) state and derive the mass action law replacing the conventional one if the IC with MC trails appears. Some structural and thermodynamic characteristics of the weak gels for all considered models are calculated and compared in Section 5.

## 2. The Lifshitz approach and Flory-like density functional description.

*2.1. The model and the Lifshitz approach.* For definiteness, we consider in this paper weak gel consisting of identical *f*-functional monomers $A_f$ , i.e. particles with $f \geq 3$ identical chemical groups *A* capable of forming A-A bonds via the reversible chemical reaction

$$A + A \leftrightarrow A_2 \qquad (2.1a)$$

The densities $\rho_1$ and $\rho_2$ of the free and bonded groups *A*, respectively, obey the mass action law (MAL)[30]

$$\rho_2 = k(T)\rho_1^2 \qquad (2.1b)$$

where $k(T)$ is the reaction equilibrium constant. The free energy *F* and partition function *Z* of this system may be written in the conventional manner:

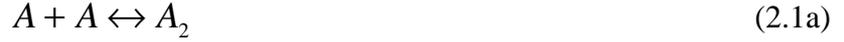

$$F(V,T,N) = -T \ln Z(V,T,N), \qquad Z(V,T,N) = \int f(\mathbf{X}) d\mathbf{X} \Big/ \Big(\lambda_T^{3N} N!\Big) \qquad (2.2)$$

Here *T* is the temperature measured in the energetic units (which corresponds to setting the Boltzmann constant *k* to be equal to unity), the denominator *N!* allows for identity of the monomers, and the heat wave length $\lambda_T$ occurs after integration the Maxwell distribution over the monomer momenta. At last, $f(\mathbf{X})$, with $\mathbf{X}=(\mathbf{r}_1, \mathbf{r}_2,...\mathbf{r}_N)$ designating a point of the space ($\mathbf{r}_i$ is the co-ordinate of the *i*-th monomer), is the distribution function in the configuration space of the system assumed to be factored as follows:

$$f(\mathbf{X}) = f_0(\mathbf{X}) f_{str}(\mathbf{X}), \qquad (2.3)$$

The first factor, $f_0(\mathbf{X})$, would be the distribution function if no bonds did exist. So, it is just the

Gibbs distribution function:

$$f_0(\mathbf{X}) = \exp\left(-U(\mathbf{X})/T\right) = \exp\left(-\sum_{i \neq j} V\left(\left|\mathbf{r}_i - \mathbf{r}_j\right|\right)\Big/2T\right), \tag{2.4}$$

($U(\mathbf{X})$ and $V(r)$ are the energy of interaction between all monomers of the system and the potential of their pair interaction, respectively.) The second factor in (2.3) describes the energy change and restriction of the configuration space of the system due to appearance of chemical bonds:

$$f_{str}(\mathbf{X}) = \sum_G \prod_G g\left(\left|\mathbf{r}_i - \mathbf{r}_j\right|\right), \tag{2.5}$$

Here the function $g(r)$ describes the additional correlation between the co-ordinates of the $i$-th and $j$-th monomers due to forming an $A$-$A$ bond between them. It is related to the chemical equilibrium constant $k(T)$ for the reaction (2.1) and the bond length $a$ as follows:

$$g_0 = \int g(r)dV = k(T) \tag{2.6}$$

$$a^2 = \int r^2 g(r)dV \Big/ \int g(r)dV \tag{2.7}$$

Now, unlike the sum in (2.4), which is taken over *all* pairs of the monomers, the product in (2.5) is taken only over the pairs of those monomers that *are bonded* or, putting it in other words, over the edges of a graph $G$ characterizing the order in which the monomers are connected to each other. Then the summation in (2.5) is performed over all possible topologically different states (graphs) $G$ of the system under consideration.

Following I.M.Lifshitz,[27] we replace the monomer coordinates by their density, which enables us to represent the partition function (2.2) as an integral over all density distributions:

$$Z(V,T,N) = \int \exp\left[-F\left(T,\{\rho(\mathbf{r})\}\right)/T\right]\delta\rho(\mathbf{r}) \tag{2.8}$$

The free energy functional $F\left(T,\{\rho(\mathbf{r})\}\right)$ appearing in (2.8) is defined as the integral over those regions of the configuration space that correspond to the density distribution $\rho(\mathbf{r})$:

$$F(T, \{\rho(\mathbf{r})\}) = -T \ln\left(\left(\int f(\mathbf{X}) d\mathbf{X}\right)_{\{\rho(\mathbf{r})\}} \Big/ N!\right) = V(\{\rho(\mathbf{r})\}) - T S(\{\rho(\mathbf{r})\}) \qquad (2.9)$$

where $V(\{\rho(\mathbf{r})\})$ and $S(\{\rho(\mathbf{r})\})$ are the potential interaction energy and entropy of the system with a given monomer density profile $\rho(\mathbf{r})$. The entropy $S(\{\rho(\mathbf{r})\})$ is shown[27] to play a role of Jacobian under replacing integration in the configurational space in (2.2) by that in the Hilbert space of the density profiles implied in (2.8). Even though the latter is a rather tricky procedure, within the saddle point (mean field) approximation, to which we restrict ourselves here, it is reduced to minimization of the free energy functional:

$$F(V, T, N) = -T \ln Z(V, T, N) = \min_{\{\rho(\mathbf{r})\}} F(T, \{\rho(\mathbf{r})\}) \qquad (2.10)$$

Now, to calculate $F(T, \{\rho(\mathbf{r})\})$ we assume that the following inequalities hold:

$$a \gg r_0, \qquad \rho a^3 \gg 1. \qquad (2.11)$$

($r_0$. being a characteristic scale of the pair monomer-monomer potential $V(r)$) and split the whole volume $V$ of the system into $\nu = V/l^3$ cubic cells of size $l$ such that

$$r_0 \ll l \ll a, \quad \rho l^3 \gg 1 \qquad (2.12)$$

Then a coarse-grained density distribution is defined by fixing the numbers $n_i \gg 1$ of the particles located in the $i$-th cell ($i=1,..,\nu$) and integration in the r.h.s. of the definition (2.9) is implied over the displacements of all particles for which the numbers $n_i$ stay fixed, which can be done in two stages. First one integrates over the displacements keeping all the particles within the original cells and afterwards over those that lead to permutations of the particles between the cells. Taking into account the factorization (2.3) and the inequalities (2.12), one sees that the first stage of this integration results in averaging of the repeatedly changing at this stage factor $f_0(\mathbf{X})$ into a coarse-grained functional. In turn, the factor $f_{str}(\mathbf{X})$, that does not change much at this stage, is averaged during the second one. As a result, the total free energy of the polymer systems splits into the sum of two statistically independent addenda:

$$\mathrm{F}\left(T,\{\rho(\mathbf{r})\}\right) = -T\ln\left[\left(\int d\mathbf{X}\, f_0(\mathbf{X})f_{str}(\mathbf{X})\right)_{\{\rho(\mathbf{r})\}}\Big/ N!\right] = \mathrm{F}^*\left(T,\{\rho(\mathbf{r})\}\right) + \mathrm{F}_{str}\left(T,\{\rho(\mathbf{r})\}\right) \qquad (2.13)$$

with the structural free energy and energetic term defined as follows:

$$\mathrm{F}_{str}\left(T,\{\rho(\mathbf{r})\}\right)/T = \ln\left[\left(\int d\mathbf{X}\, f_{str}(\mathbf{X})\right)_{\{\rho(\mathbf{r})\}}\Big/ N!\right] = \ln\left[\left(\sum_G \int d\mathbf{X} \prod_G g\left(\left|\mathbf{r}_i - \mathbf{r}_j\right|\right)\right)_{\{\rho(\mathbf{r})\}}\Big/ N!\right] \qquad (2.14)$$

$$\mathrm{F}^*\left(T,\{\rho(\mathbf{r})\}\right) = -T\ln\left[\left(\int d\mathbf{X}\, f_0(\mathbf{X})\right)_{\{\rho(\mathbf{r})\}}\Big/\left(\int d\mathbf{X}\right)_{\{\rho(\mathbf{r})\}}\right] = \mathrm{F}_0\left(T,\{\rho(\mathbf{r})\}\right) + T\mathrm{S}_{id}\left(\{\rho(\mathbf{r})\}\right) \qquad (2.15)$$

where $\mathrm{F}_0$ is the free energy of a simple liquid (the system of broken units) whose particles interact via the same potentials as the monomers do:

$$\mathrm{F}_0\left(T,\{\rho(\mathbf{r})\}\right) = -T\ln\left(\int d\mathbf{X}\, f_0(\mathbf{X})\right)_{\{\rho(\mathbf{r})\}}\Big/\left(\lambda_T^{3N} N!\right) \approx \int f_0(\rho(\mathbf{r}),T)\, dV \qquad (2.16)$$

and $\mathrm{S}_{id}$ is just the entropy of the ideal gas of point-like particles[30]:

$$\begin{aligned}
\mathrm{S}_{id}\left(\{\rho(\mathbf{r})\}\right) &= -T^{-1}\mathrm{F}_{id}\left(\{\rho(\mathbf{r})\}\right) \\
&= \frac{\left(\int d\mathbf{X}\right)_{\{\rho(\mathbf{r})\}}}{\left(\lambda_T^{3N} N!\right)} = \prod_{i=1}^{i=\nu}\left[\left(\frac{l^3}{\lambda_T^3}\frac{Z}{f!}\right)^{n_i}\frac{1}{n_i!}\right] = \int \rho(\mathbf{r})\ln\left(\frac{e\,Z}{\lambda_T^3\, f!\,\rho(\mathbf{r})}\right)dV
\end{aligned} \qquad (2.17)$$

The origin of the factor $Z/f!$ in (2.17) is due to the fact that the monomers could be considered as the point-like particles only approximately. Indeed, because of having $f$ functional groups (FG) $A$, the monomers possess some extra internal degrees of freedom to be integrated over when calculating the total partition function. It is this integrating which results in appearance of the factor $Z$ that should be divided by the symmetry index $f!$ to allow for identity of all the FG. In what follows we assume that the contribution of the internal degrees of freedom stays the same when one or more of the groups become involved into the bonds and, therefore, skip the factor $Z$. (More precisely, the corresponding changes are supposed to be additive and included into the bond function $g(r)$ definition.)

Of course, the presented consideration by itself rather explains than proves the additivity approximation (2.13) that comes back to Flory[2] and is now generally accepted in polymer theory (e.g., it is used explicitly in refs 18-21). But it is given here to elucidate, on a familiar example, the ideas whose generalization will enable us to calculate the structural free energy $F_{str}$ describing the SGT and to find its singularities in what follows.

*2.2. The density functional description of the bond formation.* Let us, similarly to the Lifshitz considerations above, split the whole volume $V$ of the system into $v = V/l^3$ cubic cells of size $l$ obeying the inequalities

$$f\rho \gg l^{-3} \gg a^{-3} \tag{2.18}$$

and define a coarse-grained density distribution by fixing the numbers $n_i \gg 1$ of the particles located in the $i$-th cell (i=1,.., $v$). Then the structural free energy (2.14) corresponding to this density distribution can be calculated as follows.

First, we specify for every cell a number $n_i^A$ of the "active" FG $A$ that do participate in formation of bonds. In terms of the density functional description it corresponds to fixing the distributions $\rho(\mathbf{r}) = n_i/l^3$, $\rho_A(\mathbf{r}) = n_i^A/l^3$ with $\mathbf{r}$ corresponding to the center of the $i$-th cell. The distributions $\{\rho_A(\mathbf{r})\}$, $\{\rho(\mathbf{r})\}$ characterize a macroscopic state of the system. Now, there are many microscopic states with different partitioning of $N = \sum n_i$ monomers between the cells with the occupation numbers $n_i$ and $fn_i$ of the FG located in the $i$-th cell into $n_i^A$ reacted (active) and $fn_i - n_i^A$ unreacted ones. Besides, the active groups may be connected by labile bonds in many ways (graphs $G$) for every macroscopic state. So, we can write the desired free energy $\mathbf{F}_{str}$ of the fully equilibrium system with fixed $\{\rho(\mathbf{r})\}$ in terms of the sum over all its partially equilibrium states with fixed $\{\rho_A(\mathbf{r})\}, \{\rho(\mathbf{r})\}$:

$$\begin{aligned}
F_{str}(\{\rho(\mathbf{r})\}) &= -T \ln \int \delta \rho_A(\mathbf{r}) \exp\left[-F_{str}(\{\rho(\mathbf{r})\}, \{\rho_A(\mathbf{r})\})/T\right] \\
&= \min_{\{\rho_A(\mathbf{r})\}} F_{str}(\{\rho(\mathbf{r})\}, \{\rho_A(\mathbf{r})\})
\end{aligned} \tag{2.19}$$

where $F_{str}(\{\rho(\mathbf{r})\}, \{\rho_A(\mathbf{r})\})$ is the free energy of the partially equilibrium (with respect to all partitions of the reacted groups and all ways to connect them into clusters as described by graphs $G$) macroscopic state. The last equality in (2.19) corresponds to the mean field approximation. In turn, for the free energy $F_{str}(\{\rho(\mathbf{r})\},\{\rho_A(\mathbf{r})\})$ of the partially equilibrium state with $\{\rho_A(\mathbf{r})\},\{\rho(\mathbf{r})\}$ also holds an additivity approximation

$$F_{str}(\{\rho(\mathbf{r})\},\{\rho_A(\mathbf{r})\}) = F_{part}(\{\rho(\mathbf{r})\},\{\rho_A(\mathbf{r})\}) + F_{clust}(\{\rho_A(\mathbf{r})\}) \qquad (2.20)$$

This additivity is due to the fact that two averagings corresponding to summing over different partitions, resulting into the first term in (2.20), and different ways to bond the active groups *A,* resulting into the second one, are factored due to inequality (2.18). Now we proceed to explicit calculation of these terms.

*2.2.1. The partitioning free energy and the first appearance of the broken monomer identity (BMI) concept.* We calculate here the term $F_{part}(\{\rho(\mathbf{r})\},\{\rho_A(\mathbf{r})\})$ in two equivalent ways in order to enable a reader to get the first clear idea of the BMI concept.

A. First, we find the number of all possible ways to choose for every cell $n_i^A$ *active* groups among all $fn_i$ groups present in this cell. This number may be written [30] as follows:

$$W_{choice} = \prod_{i=1}^{i=v} \frac{(fn_i)!}{(fn_i - n_i^A)!n_i^A!} = \exp\left\{-\frac{F_{choice}(\{\rho(\mathbf{r})\},\{\rho_A(\mathbf{r})\})}{T}\right\}$$
$$F_{choice}(\{\rho(\mathbf{r})\},\{\rho_A(\mathbf{r})\}) = f\int dV\rho(\mathbf{r})[\Gamma(\mathbf{r})\ln\Gamma(\mathbf{r}) + (1-\Gamma(\mathbf{r}))\ln(1-\Gamma(\mathbf{r}))] \qquad (2.21)$$

where the local conversion is defined as the ratio of the local reacted and all groups densities:

$$\Gamma(\mathbf{r}) = \rho_A(\mathbf{r})/(f\rho(\mathbf{r})) \qquad (2.22)$$

Next, we still are to allow for the conventional translational entropy of the monomers to which end we have to integrate over their co-ordinates inside every cell. As a result, we get the following final expression for the partitioning free energy:

$$F_{part}(\{\rho(\mathbf{r})\},\{\rho_A(\mathbf{r})\}) = F_{choice}(\{\rho(\mathbf{r})\},\{\rho_A(\mathbf{r})\}) - T\ln\left(\left(\int d\mathbf{X}\right)_{\{\rho(\mathbf{r})\}}\Big/N!\right)$$
$$= F_{choice}(\{\rho(\mathbf{r})\},\{\rho_A(\mathbf{r})\}) - T\,S_{id}(\{\rho(\mathbf{r})\}) \qquad (2.23)$$

Substituting into (2.23) the formulas (2.17), (2.21) gives the desired explicit expression for the partitioning free energy $F_{part}$:

$$F_{part}\big(\{\rho(\mathbf{r})\},\{\rho_A(\mathbf{r})\}\big)=f\int dV\rho(\mathbf{r})\big[\Gamma(\mathbf{r})\ln\Gamma(\mathbf{r})+\big(1-\Gamma(\mathbf{r})\big)\ln\big(1-\Gamma(\mathbf{r})\big)\big]$$
$$+\int\rho(\mathbf{r})\ln\big(\lambda_T^3\ f!\,\rho(\mathbf{r})/e\big)dV \qquad (2.24)$$

Thus, in (2.23) $F_{part}$ is defined as the sum of the translational free energy of *all* monomers, without any specifying of their structure, and that of the redistribution of *all* FG into the reacted and unreacted ones (without specifying to which monomers these groups do belong).

B. On the other hand, any fixed distribution of the reacted FG induces a splitting of all monomers into classes with respect to the number $i$ ($0{\le}i{\le}f$) of the reacted FG the monomers have. Such a splitting is characterized by the coarse-grained densities $\rho_i(\mathbf{r})$ of the monomers with $i$ reacted groups. The monomers with different values of $i$ cannot be considered as identical anymore. Thus, the total translational free energy for a specified distribution $\{\rho_i(\mathbf{r})\}$ can be written down as the sum of the translational free energies for all classes of monomers:

$$\widetilde{F}_{part}\big(\{\rho_i(\mathbf{r})\}\big)=T\sum_{i=0}^{i=f}\big(\rho_i(\mathbf{r})\ln\big[\rho_i(\mathbf{r})\lambda_T^3\,(f-i)!i!/e\big]\big) \qquad (2.25)$$

where the symmetry index $i!(f\text{-}i)!$ allows for identity of all $i$ the reacted and $f\text{-}i$ unreacted groups and distinction between the reacted and unreacted groups. The fact that *the symmetry indices of the monomers belonging to different classes are different* suggests treating them as different particles and considering the monomer distribution over such classes as *a breaking of the monomer identity*. For specified total densities of all monomers and all reacted groups

$$\sum_{i=0}^{i=f}\rho_i(\mathbf{r})=\rho(\mathbf{r}),\quad\sum_{i=0}^{i=f}i\rho_i(\mathbf{r})=\rho_A(\mathbf{r}) \qquad (2.26)$$

the equilibrium distribution $\{\bar{\rho}_i(\mathbf{r})\}$ is expected to minimize the free energy (2.25) under the additional conditions (3.2) thus giving the desired value of the partitioning free energy:

$$F_{part}\big(\{\rho(\mathbf{r})\},\{\rho_A(\mathbf{r})\}\big)=$$
$$\min\left\{\widetilde{F}_{part}\big(\{\rho_i(\mathbf{r})\}\big)+\int\left(\mu(\mathbf{r})\left[\tilde{n}(\mathbf{r})-\sum_{i=0}^{f}\rho_i(\mathbf{r})\right]+\lambda(\mathbf{r})\left[\rho_A(\mathbf{r})-\sum_{i=0}^{f}i\rho_i(\mathbf{r})\right]\right)dV\right\} \qquad (2.27)$$

Performing such a minimization gives

$$\rho_i(\mathbf{r}) = z(\mathbf{r})\left[\Phi(\mathbf{r})\right]^i \Big/ (i!(f-i)!), \quad z(\mathbf{r}) = \lambda_T^{-3}\exp\mu(\mathbf{r}), \quad \Phi(\mathbf{r}) = \exp\lambda(\mathbf{r}) \qquad (2.28)$$

Substituting (2.28) into conditions (2.26) we get

$$\rho(\mathbf{r}) = \sum_{i=0}^{i=f}\rho_i(\mathbf{r}) = z(\mathbf{r})\left(1+\Phi(\mathbf{r})\right)^f \Big/ f!, \quad \rho_A(\mathbf{r}) = \sum_{i=0}^{i=f}i\,\rho_i(\mathbf{r}) = z(\mathbf{r})\,\Phi(\mathbf{r})\left(1+\Phi(\mathbf{r})\right)^{f-1}\Big/(f-1)!$$

which enables us to express the parameters $z(\mathbf{r}),\Phi(\mathbf{r})$ related to the Lagrange multipliers $\mu(\mathbf{r})$, $\lambda(\mathbf{r})$ in terms of the local monomer conversion and total density:

$$\Gamma(\mathbf{r}) = \rho_A(\mathbf{r})\big/(f\rho(\mathbf{r})) = \Phi(\mathbf{r})\big/(1+\Phi(\mathbf{r})) \to \Phi(\mathbf{r}) = \Gamma(\mathbf{r})\big/(1-\Gamma(\mathbf{r})),$$
$$\lambda(\mathbf{r}) = \ln\left[\Gamma(\mathbf{r})\big/(1-\Gamma(\mathbf{r}))\right], \quad z(\mathbf{r}) = f!\rho(\mathbf{r})(1-\Gamma(\mathbf{r}))^f, \quad \mu(\mathbf{r}) = \ln\left(\rho\lambda_T^3 f!(1-\Gamma)^f\right) \qquad (2.29)$$

One can check readily by substituting (2.28), (2.29) into (2.27) that the presented procedure results precisely in the expression (2.24) for $F_{part}$ and gives, in addition, the values of the equilibrium densities of the monomers having exactly $i$ reacted groups:

$$\rho_i(\mathbf{r}) = \frac{\rho(\mathbf{r})f!}{i!(f-i)!}\left(\Gamma(\mathbf{r})\right)^i\left(1-\Gamma(\mathbf{r})\right)^{f-i}, \quad \Gamma(\mathbf{r}) = \frac{\rho_A(\mathbf{r})}{f\rho(\mathbf{r})} \qquad (2.28a)$$

Thus, the BMI concept gives an alternative way to treat the free energy $F_{part}$ as the sum of the translational free energies of the monomers partitioned over different topological classes as well as to find the equilibrium distribution over such classes. In this way the account of the redistribution of *all* FG into the reacted and unreacted ones is replaced by account of different symmetry indices for different classes. Therewith, the concentration dependence of the partitioning (2.28) (or, putting it in other words, the monomer identity breaking) is gradual.

*2.2.2. The free energy of the bond formation.* After a microscopic state is fixed by the numbers and locations of the active groups in each cell, the free energy $F_{clust}$ and entropy $S_{clust}$ corresponding to all possible ways to connect the active groups by bonds are as follows:

$$F_{clust}\left(\{\rho_A(\mathbf{r})\}\right) = -T\,S_{clust}\left(\{\rho_A(\mathbf{r})\}\right), \quad S_{clust}\left(\{\rho_A(\mathbf{r})\}\right) = \ln\left(\sum_G\prod_G g\left(\left|\mathbf{r}_i - \mathbf{r}_j\right|\right)\right)_{\{\rho_A(\mathbf{r})\}} \qquad (2.30)$$

It is the contribution $F_{clust}$ that contains the most important information about the cluster formation and structure and, in particular, about existence or absence of any singularities in the weak gel behavior. To provide necessary understanding of this contribution we calculate $F_{clust}$ in every section from a new viewpoint, thus gradually revealing more and more complicated structure of the IC of the reversible bonds.

First we calculate the functional $F_{clust}$ without invoking any idea of the IC structure. Indeed, the presented considerations make sense also for a model system of $N$ dimers formed in the volume $V$ via complete association of $2N$ $1$-functional monomers. For this system $\rho_A(\mathbf{r}) \equiv \rho(\mathbf{r})$ so that $\Gamma(\mathbf{r}) \equiv 1$ and the free energy of choice (2.21) equals zero. Thus, we get

$$F_2\left(\{\rho(\mathbf{r})\}\right) = -T\,S_2\left(\{\rho(\mathbf{r})\}\right) = -T\,S_{bond}\left(\{\rho(\mathbf{r})\}\right) + F_{id}\left(\{\rho(\mathbf{r})\}\right) \tag{2.31a}$$

$$S_{bond}\left(\{\rho_A(\mathbf{r})\}\right) = \ln\left(\sum_{\substack{bonds \\ distributions}} \prod_{bonds} g\left(\left|\mathbf{r}_i - \mathbf{r}_j\right|\right)\right)_{\{\rho_A(\mathbf{r})\}} \tag{2.31b}$$

where $S_{bond}$ is the entropy of the distribution of $N$ bonds between $2N$ groups $A$ as a functional of the density profile $\rho_A(\mathbf{r})$ of these groups. On the other hand, the entropy $S_2(\{\rho(\mathbf{r})\})$ of the fully associated $1$-functional monomers as calculated straightforwardly by Erukhimovich[25] is:

$$S_2\left(\{\rho(\mathbf{r})\}\right) = (1/2)\int dV\,\rho(\mathbf{r})\ln\left((e/\rho(\mathbf{r}))\left((\hat{g}\psi)(\mathbf{r})/\psi(\mathbf{r})\right)\right), \tag{2.32a}$$

where the function $\psi(\mathbf{r})$ is implicitly determined by the relationship

$$\rho(\mathbf{r}) = \mathcal{N}\,\psi(\mathbf{r})\hat{g}\psi \tag{2.32b}$$

$\mathcal{N}$ being an irrelevant normalization constant and the operator $\hat{g}\psi$ is defined in Appendix. It follows from (2.17), (2.31), (2.32) the desired expression for the quantity $S_{bond}$ :

$$S_{bond}\left(\{\rho_A(\mathbf{r})\}\right) = S_2\left(\{\rho_A(\mathbf{r})\}\right) - S_{id}\left(\{\rho_A(\mathbf{r})\}\right) = \int \frac{\rho_A(\mathbf{r})}{2}\ln\left(\frac{\rho_A(\mathbf{r})}{e}\frac{(\hat{g}\psi)(\mathbf{r})}{\psi(\mathbf{r})}\right)dV \tag{2.33}$$

Let now apply the same considerations to a model system of $N$ bifunctional monomers $A_2$ forming, via complete association of the FG $A$, one polymer chain filling the volume $V$. (A finite fraction of cycled chains existing in such a system is known to be a small quantity of the order $O(1/(f\rho a^3))$. So, it is neglected in our approximation assuming that the inequality (2.18) holds.) The structural free energy and entropy of such a system are

$$F_\infty(\{\rho(\mathbf{r})\}) = -T\,S_\infty(\{\rho(\mathbf{r})\}) = -T\,S_{bond}^{(\infty)}(\{\rho(\mathbf{r})\}) + F_{id}(\{\rho(\mathbf{r})\}) \tag{2.34a}$$

where

$$S_{bond}^{(\infty)}(\{\rho(\mathbf{r})\}) = \ln\left( \sum_{\substack{\text{all infinite chains connecting} \\ \text{the points } \mathbf{r_1},...,\mathbf{r}_N}} \prod_{\substack{\text{an infinite chain} \\ \text{via points } \mathbf{r_1},...,\mathbf{r}_N}} g\left(\left|\mathbf{r}_i - \mathbf{r}_j\right|\right) \right)_{\{\rho_A(\mathbf{r})\}} \tag{2.34b}$$

is the entropy of the distribution of $N$ bonds between $N$ monomers $A_2$ as a functional of the density profile $\rho(\mathbf{r})$ of the monomers.

On the other hand, the entropy $S_\infty(\{\rho(\mathbf{r})\})$ corresponds to the number of all ways to locate $N$ monomers $A_2$ in the volume $V$ and connect them with bonds A-A into one (infinite in the thermodynamic limit $N,V\rightarrow\infty$, $\rho=N/V=const$) chain. Putting it in other words, it is nothing but the entropy corresponding to the number of all ways to fill the space with such infinite chain whose monomers density profile is $\rho(\mathbf{r})$. The expression for the latter entropy, found by I.M.Lifshitz,[27] reads:

$$S_\infty(\{\rho(\mathbf{r})\}) = \int dV\,\rho(\mathbf{r})\ln\left(\hat{g}\psi/\psi\right) \tag{2.35}$$

where the functions $\rho(\mathbf{r})$ and $\psi(\mathbf{r})$ also obey the relationship (2.32b). The bonding entropy $S_{bond}^{(\infty)}(\{\rho(\mathbf{r})\})$ found from (2.17),(2.34),(2.35) is

$$\begin{aligned} S_{bond}^{(\infty)}(\{\rho(\mathbf{r})\}) &= \left(S_\infty(\{\rho(\mathbf{r})\}) - S_{id}(\{\rho(\mathbf{r})\})\right) = \int dV\,\rho(\mathbf{r})\ln\left(2\rho(\mathbf{r})(\hat{g}\psi)(\mathbf{r})/[e\psi(\mathbf{r})]\right) \\ &= (1/2)\int dV\,\rho(\mathbf{r})\ln\left((\rho(\mathbf{r})/e)((\hat{g}\psi)(\mathbf{r})/\psi(\mathbf{r}))\right) \end{aligned} \tag{2.36}$$

Comparison of the expressions (2.33) and (2.36) shows that the bonding entropy is the same, within the coarse-grain density functional description, either when the groups A are separated, as in (2.33), or coupled into monomers $A_2$, as in (2.36). We conclude that the entropy of the distribution of $N$ bonds between $2N$ groups A is a functional of the total group A density $\rho_A(\mathbf{r})$ only and does not depend on the distribution of the groups between monomers. For uniform monomer and reacted group densities distributions ($\rho(\mathbf{r})=\rho$, $\rho_A(\mathbf{r})=\rho_A$) this entropy reads:

$$S_{bond}\left(\rho_A(\mathbf{r})\right)\Big|_{\rho_A(\mathbf{r})=\rho_A=const} = V S_{bond}(\rho_A), \quad S_{bond}(\rho_A) = \rho_A \ln\left(g_0 \rho_A/e\right)/2 \qquad (2.37)$$

(Here and henceforth we designate the specific (per unit volume) quantities that are some functions of the average densities by italic unlike the corresponding functionals including an integration over the whole volume $V$ of the system we have used until now and designated by the same straight letters.)

*2.3. Flory-like density functional description of the weak gels.* Now, using (2.17), (2.20), (2.24) and (2.37), we can write the expression (2.20) for the specific virtual structural free energy of the system of $f$-functional monomers $A_f$ as follows:

$$F_{str}(\rho,\rho_A) = F_{part}(\rho,\rho_A) - T S_{id}(\rho) = T\{f\rho[\Gamma \ln \Gamma + (1-\Gamma)\ln(1-\Gamma)]$$
$$+ \rho \ln\left(\lambda_T^3 f!\rho/e\right) - (\rho_A/2)\ln\left(g_0\rho_A/e\right)\} \qquad (2.38)$$

Taking into account the definition (2.22) of the conversion $\Gamma$ and minimizing the r.h.s. of the expression (2.38) with respect to $\rho_A$, as prescribed in (2.19), we get the MAL:

$$\Gamma\big/(1-\Gamma)^2 = \rho_{tot} = f g_0 \rho \qquad (2.39)$$

and the expression for the specific structural free energy:

$$F_{str}(\rho,T) = V \min_{\{\rho_A\}} F_{str}(\rho,\rho_A,T) = V F_{id}(\rho,T) + f N F_{Flory}(\tilde{\rho}_{tot},T) \qquad (2.40)$$

We introduced here the reduced density $\tilde{\rho}_{tot}$ of all FG and bonding free energy per one FG

$$F_{Flory}(\tilde{\rho}_{tot},T) = T\left[(\Gamma/2) + \ln(1-\Gamma)\right] \qquad (2.40a)$$

where the function $\Gamma(\tilde{\rho}_{tot})$ is determined by the MAL (2.39).

At last, the expression for the weak gel pressure follows after substituting (2.40) into (2.13) and differentiating over the volume:

$$P(\rho,T) = P_0(\rho,T) - f \rho \Gamma/2 = P_0(\rho,T) - T \nu_{bond} \qquad (2.41)$$

where $\nu_{bond}$ is the number of bonds per unit volume and $P_0(\rho,T)$ is the pressure of the system of broken units. Eq 2.41 has very clear physical meaning: formation of every new bond decreases the number of the translational degrees of freedom by unity. As to our knowledge, the expressions (2.40), (2.41) were presented first in refs 26 and 31, respectively, under studying crosslinking of polymer chains and generalized[12] for the system of $A_f$-monomers with an arbitrary $f \geq 3$. Later they were rederived[18-21] and related to the Flory approach.[21,25]

We see that the course-grained density functional description[12,25] reviewed in this section leads precisely to the later results of Semenov and Rubinstein.[21] However, the expressions (2.40), (2.41) that determine the thermodynamic behavior of the weak gels give no idea of the gel structure for different values of the conversion $\Gamma$. The only thermodynamic indication of the sol-gel transition in these systems is that for the ideal weak gels, where $P_0(\rho,T)=T\rho$, the compressibility calculated with due regard for the MAL (2.39) reads:

$$T\left(\partial \rho / \partial P_{\mathrm{id}}^{\mathrm{gel}}\right)_T = (1+\Gamma)/[1-(f-1)\Gamma] \qquad (2.42)$$

Thus, the ideal weak gels are thermodynamically unstable if the Flory-Stockmayer condition

$$\Gamma > \Gamma_c = (f-1)^{-1} \qquad (2.43)$$

holds ($\Gamma_c$ is the value of the conversion at the sol-gel threshold).

But even this indication disappears if a potential interaction between the monomers is included (such an interaction would modify the $\rho$-dependence of the pressure $P_0(\rho,T)$ and, accordingly, shift the value of the critical (as to instability) conversion $\Gamma$ from $\Gamma_c$). Thus, to be able to distinguish the sol and gel phases as well as determine their structures, one needs a more detailed and sophisticated description we present in the next section via the broken monomer identity (BMI) concept.

## 3. The BMI concept and the structured density functional description of the gel fraction in Stockmayer and Flory models.

*3.1. The description of the gel fraction structure.* When we first applied the BMI concept to calculate $F_{part}$ in subsection 2.2.1, the partitioning of the monomers into different topological classes was quite trivial. Now, to step further we address a more complicated problem: how is it possible to find, within the coarse-grained density functional description, any difference between the monomers belonging to the sol and gel fractions, respectively?

A nontrivial procedure to implement this possibility is as follows. We choose a finite cell (called further window) of a size $L$ and scan all the clusters of chemical bonds A-A that are located within the window. If such a cluster is completely confined to the window, it belongs, evidently, - just by definition, - to the sol-fraction, and we color all the bonds, belonging to such a finite cluster, red. In the opposite case some of the chemical bonds overstep the window boundaries. If there is a connected trail of the bonds linking two "overstepping" bonds, we refer to all the bonds belonging to such a trail as bilateral and color them blue. The rest of the bonds located in the window (neither red nor blue) belong to the trails terminated by an unreacted group (dead end) from one side and a blue trail of bonds or boundary of the window from another one. We refer to these bonds as unilateral, color them yellow and, besides, amend them by arrows directed towards dead ends (see Figure 2). Now, let the size $L$ of the window be increasing. Then, in the larger windows, some of the initially blue and yellow bonds may turn out to belong to a finite cluster; so, they should be recolored red. Thus, as $L$ increases, the amount of the bi- and unilateral (blue and yellow) bonds decreases. If it reaches zero, which means that only red (sol) bonds remain in the limit $L \rightarrow \infty$, we say that the system is in the sol phase. In the opposite case, when the fraction of the blue and yellow bonds approaches a finite value even in the limit $L \rightarrow \infty$, one can not help but refer to these bonds as those belonging to the infinite cluster (IC). Then we say that the system is in the gel phase, the finite clusters of red bonds forming sol fraction and all other ones belonging to gel fraction. (For the clusters of labile bonds the described coloring procedure should be done using instant snapshots of the clusters.) Following Flory and Stockmayer, we assume also in this section that, however large the size $L$ of the window is, any cluster confined to the window has a tree-like structure and does not contain any cycles (closed trails of the bonds).

Now, all the monomers belonging to finite clusters of the sol-fraction can be considered as identical in the sense that all of them are involved into finite trails of bonds only. The monomers that belong to sol-fraction and their FG are further called $S_f$-monomers and $S$-groups, respectively. On the contrary, the monomers and FG that belong to the IC can be further differentiated.

The FG that form a blue bond are, evidently, indistinguishable. We call them $I$-groups to indicate that they generate the infinite trails. On the contrary, the FG that form a yellow bond differ both from the $I$-groups and each other. The groups, that are starting points of the trails terminated by dead ends, and those terminated by blue trails will be called $D$- and $D^+$-groups respectively (see incoming and outgoing yellow arrows in Figure 3b,c). The unreacted FG are included to the D-groups. Thus, all the $D^+$-groups are involved in the bonds just by definition and thus had reacted, the numbers of the $D^+$-groups and the $D$-groups forming the yellow bonds $D$-$D^+$ being the same, whereas a part of the $D$-groups stays unreacted. To describe this fact we introduce the conversion $\Gamma_D$ of the $D$-groups as follows:

$$\rho_D \Gamma_D = \rho_{D^+} \tag{3.1}$$

where the densities $\rho_D, \rho_{D^+}$ of the corresponding groups are introduced.

Thus, we defined three topological types of the FG ($I$, $D$ and $D^+$) and two types of chemical bonds (blue $I$-$I$ and yellow $D$-$D^+$) that belong to the IC. Accordingly, there are only the following topological types of the IC monomers (see Figure 3): 1) the "dangling" monomers $D^+D_{f-1}$ forming a fringe hanging from the IC backbone (all the bonds adjacent to such monomers are yellow), 2) the backbone monomers $I_2D_{f-2}$; and 3) the junction monomers $I_nD_{f-n}$ with $3 \leq n \leq f$ (exactly $n$ of the bonds adjacent to such monomers are blue). Thus, the quantitative description of the gel fraction structure is provided by the distribution of the average numbers per unit volume $\rho_1$ of the dangling monomers, $\rho_2$ of backbone ones with 2 blue bonds and $\rho_n$ of the junction monomers with $n$ blue bonds.

Now, before to proceed to quantitative treatment of the classic Flory and Stockmayer gel models, we notice that the presented definitions of the monomer and group densities entail the following relationships between them:

$$\rho_{D^+} = \rho_1, \quad \rho_D = \sum_{i=1}^{i=f} (f-i)\rho_i, \quad \rho_g = \sum_{i=1}^{f} \rho_i, \quad \rho_I = \sum_{i=2}^{f} i\rho_i. \tag{3.2}$$

where $\rho_g$ and $\rho_I$ are, respectively, the density of all monomers belonging to the gel-fraction and $I$-groups.

It is also useful to calculate the conversion $\Gamma_{IC}$ of the IC in terms of the distribution $\{\rho_i\}$. To this end we notice that the density of all reacted groups belonging to the gel-fraction is:

$$\rho_g^A = \rho_I + \rho_{D^+} + \rho_D \Gamma_D = 2\rho_1 + \sum_2^f i\rho_i \tag{3.3}$$

Therefore, the desired conversion $\Gamma_{IC}$ can be written as follows

$$\Gamma_{IC} = \rho_g^A \Big/ \big(f\,\rho_g\big) = (2/f) + \sum_3^f (i-2)\rho_i \Big/ \big(f\,\rho_g\big) \tag{3.4}$$

On the other hand, the conversion of a tree-like cluster consisting of $n$ monomers and, thus, ($n$-$1$) bonds is $\Gamma_n = 2(n-1)/(fn)$. Therefore, introducing $\Gamma_\infty^{tree} = \lim_{n\to\infty} \Gamma_n^{tree} = 2/f$ one can rewrite (3.4) as follows:

$$\Gamma_{IC} = \Gamma_\infty^{tree} + \sum_3^f (i-2)\rho_i \Big/ \left( f \sum_1^f \rho_i \right), \tag{3.5}$$

3.2. *The structured density functional description for the Stockmayer-Tanaka (ST) gel model.* It follows from (3.5) that in the absence of the junction monomers ($\rho_i = 0$ for $i > 2$) the conversion of the IC would equal precisely the value of that for the infinite tree-like cluster and thus fit the Stockmayer conjecture $\Gamma_{IC} = \Gamma_\infty^{tree}$. Thus, it is natural to identify the ST gel model with the assumption that the only topological types of the monomers present in the weak gels are sol and dangling ones as well as the backbone monomers $I_2D_{f-2}$. Similarly to the treatment in subsection *2.2.3,* the desired structural free energy $F_{str}(\rho)$ of such a system reads:

$$\mathrm{F}_{str}(\rho)/V = \min F_{str}(\rho,\rho_g), \quad F_{str}(\rho,\rho_g) = F_{str}^s(\rho_s) + F_{str}^g(\rho_g) \tag{3.6}$$

where the total densities $\rho_s$ and $\rho_g$ of the monomers belonging to the sol- and gel-fractions, respectively, satisfy condition of the fixed total density of the weak gel:

$$\rho_s + \rho_g = \rho \tag{3.7}$$

the desired equilibrium value of the gel-fraction density $\rho_g$ being found as the root of the corresponding extremal equation:

$$\frac{\partial F_{str}(\rho,\rho_g)}{\partial \rho_g} = \frac{\partial F_{str}^g(\rho_g)}{\partial \rho_g} - \frac{\partial F_{str}^s(\rho_s)}{\partial \rho_s}\bigg|_{\rho_s=\rho-\rho_g} = \mu_{str}^g - \mu_{str}^s = 0 \tag{3.8}$$

where the structural contributions into chemical potentials of the monomers pertaining to sol and gel fractions are introduced. As consistent with (2.37), we have for the structural free energy of the sol-fraction in (3.6):

$$F_{str}^s(\rho_s) = \min F_{str}^s(\rho_s,\rho_s^A), \tag{3.9}$$

where $\rho_s^A = f\rho\,\Gamma_s$,

$$\frac{F_{str}^s(\rho_s,\rho_s^A)}{T} = \rho_s \ln \frac{f!\rho_s\lambda_T^3}{e} + f\rho_s\left(\Gamma_s \ln \Gamma_s + (1-\Gamma_s)\ln(1-\Gamma_s)\right) - \frac{\rho_s^A}{2}\ln\frac{\rho_s^A g_0}{e} \tag{3.10}$$

and minimization in (3.9) is to be done with respect to the density $\rho_s^A$ of the reacted *S*-groups or, which is the same, the sol conversion $\Gamma_s$. As a result we get:

$$\Gamma_s/(1-\Gamma_s)^2 = \tilde{\rho}_{\mathrm{tot}}^s = fg_0\rho_s, \tag{3.11a}$$

$$F_{str}^s(\rho_s)/T = \rho_s \ln\left[f!\rho_s\lambda_T^3/e\right] + g_0^{-1}\mathrm{F}_{\mathrm{Flory}}(\tilde{\rho}_{\mathrm{tot}}^s), \tag{3.11b}$$

where the function $\mathrm{F}_{\mathrm{Flory}}$ is determined by expression (2.40a), and

$$\begin{aligned}\partial F_{str}^s(\rho_s)/\partial\rho_s = \mu_{str}^s &= T\left\{\ln\Gamma_s + (f-2)\ln(1-\Gamma_s) + \ln\left[(f-1)!\lambda_T^3/g_0\right]\right\} \\ &= \mu_{id}^s(\rho_s) + fT\ln(1-\Gamma_s)\end{aligned} \tag{3.11c}$$

In turn, the structural free energy of the gel-fraction in (3.6) is defined as follows:

$$F_{str}^g(\rho_g) = \min F_{str}^g(\rho_1,\rho_2,\Gamma_D) \tag{3.12}$$

where

$$F_{str}^g(\rho_1, \rho_2, \Gamma_D)/T = \rho_1 \ln \frac{(f-1)!\rho_1 \lambda_T^3}{e} + \rho_2 \ln \frac{2(f-2)!\rho_2 \lambda_T^3}{e}$$
$$- \rho_1 \ln \frac{\rho_1 g_0}{e} - \rho_2 \ln \frac{2\rho_2 g_0}{e} + \rho_D(\Gamma_D \ln \Gamma_D + (1-\Gamma_D)\ln(1-\Gamma_D)) \quad (3.13)$$

The first two terms in (3.13) are the free energies of the ideal gases for the dangling and backbone monomers with due regard for their symmetry indices. The next two terms are the free energies of forming the directed yellow bonds and symmetric blue ones given by (A9) and (2.36), respectively. The last term is the free energy of choice of the active $D$-groups between all of them (all $D^+$- and $I$-groups are active by definition thus giving no extra free energy of their choice). Minimization in (3.13) is to be done with respect to the dangling groups' conversion $\Gamma_D$ and both densities of the dangling and backbone monomers under the additional conditions (3.2).

Now, rewriting one of the relationships (3.2) for the considered Stockmayer model in the form

$$\rho_D = (f-1)\rho_1 + (f-2)\rho_2 \quad (3.14)$$

and substituting (3.2), (3.14) into (3.1) we get the expression for the $D$-groups conversion:

$$\Gamma_D = [f - 1 + (\rho_2/\rho_1)(f-2)]^{-1} \quad (3.15)$$

Inverting (3.15) gives the relevant densities:

$$\rho_1 = \rho_g(f-2)\Gamma_D/(1-\Gamma_D), \quad \rho_2 = \rho_g[1-(f-1)\Gamma_D]/(1-\Gamma_D),$$
$$\rho_D = \rho_g(f-2)/(1-\Gamma_D) \quad (3.16)$$

and substituting (3.16) into (3.13) reduces the structural free energy of gel to

$$F_{str}^g(\rho_1, \rho_2, \Gamma_D)/T = \rho_g(f-2)\{[\Gamma_D/(1-\Gamma_D)]\ln[\Gamma_D(f-1)] + \ln(1-\Gamma_D)\}$$
$$+ \rho_g \ln[(f-2)!\lambda_T^3/g_0] \quad (3.17)$$

At last, minimizing (3.17) with respect to $\Gamma_D$, we get

$$\Gamma_D = (f-1)^{-1}, \quad \rho_2 = 0, \quad (3.18a)$$

$$F_{str}^g\left(\rho_g\right)/T = \rho_g\left\{(f-2)\ln\left[(f-2)/(f-1)\right] + \ln\left[(f-2)!\lambda_T^3/g_0\right]\right\} \tag{3.18b}$$

$$\mu_{str}^g\left(\rho_g\right) = \partial F_{str}^g\left(\rho_g\right)/\partial\rho_g = T\left\{(f-2)\ln\left[(f-2)/(f-1)\right] + \ln\left[(f-2)!\lambda_T^3/g_0\right]\right\} \tag{3.18c}$$

One can readily check now by substituting (3.11c) and (3.18c) into the equation of extremum (3.8) that the only solution of this equation determining the equilibrium amount of the sol fraction and the corresponding sol conversion take the following values:

$$\rho_s = \rho_c = (f-1)\Big/\left[f\,g_0(f-2)^2\right], \quad \Gamma_s^c = \Gamma_D = (f-1)^{-1}, \quad \rho_g = \rho - \rho_c. \tag{3.19}$$

which means that the state of the sol fraction stays critical and does not change with increase of the total monomer density above the sol-gel threshold (i.e. for $\rho > \rho_c$). Moreover, combining (3.11c) and (3.18c), we see that the ideal weak gel chemical potential $\mu_{str}$ monotonously increases with the total density $\rho$ increase until the sol-gel threshold and stays constant above the threshold. As consistent with (3.11c) and (3.18c), the first derivative of $\mu_{str}$ with respect to $\rho$ equals zero for $\rho = \rho_c \pm \varepsilon, \varepsilon \to 0$ thus providing continuity of this derivative either. So, it is only the second derivative of $\mu_{str}$ (and, thus, the third one of the structural free energy $F_{str}(\rho)$) that jumps when the gel-fraction appears. Therefore, within the considered approximation the SGT turns out to be the third order phase transition.

To find the final expression for the total specific structural free energy of the gel phase within the ST approximation, we substitute the expression (3.11b) for the free energy of sol fraction at sol-gel-threshold and (3.18b) for that of gel fraction into (3.6) which gives

$$F_{str}\left(\rho,T\right) = VF_{id}\left(\rho,T\right) + f\,N\begin{cases} F_{\text{Flory}}\left(\tilde{\rho}_{\text{tot}}\right), & \Gamma \le \Gamma_c = (f-1)^{-1} \\ F_{\text{Stockmayer}}\left(\tilde{\rho}_{\text{tot}}\right), & \Gamma \ge \Gamma_c \end{cases}, \tag{3.20a}$$

$$\frac{F_{\text{Stockmayer}}}{T} = \frac{1}{f}\left((f-2)\ln\frac{f-2}{f-1} + \ln\frac{e\Gamma_e}{\tilde{\rho}_{\text{tot}}}\right) - \frac{\tilde{\rho}_{\text{tot}}^{-1}}{2f(f-2)} \tag{3.20b}$$

Thus, basing on the only assumption that no junction monomers are present in the strictly tree-like thermodynamically equilibrium IC, we managed to derive all the results usually supposed to be the Stockmayer conjectures only. These results, however, are unstable with respect to including of the junction monomers into consideration.

*3.3. The structured density functional description for the Flory gel model.* Indeed, admit now that all topological types of the tree-like IC monomers (not only the dangling and backbone ones but also the junction monomers) are present in the IC in the fractions consistent with conditions of the thermodynamic equilibrium. Then, as consistent with (3.5), $\Gamma_{IC} > \Gamma_\infty^{tree}$ which is quite natural: it is just in the junction monomers where the infinite trails coming from infinity meet thus forming a cycle.

The desired structural free energy $F_{str}(\rho)$ of such a system can be written as follows:

$$F_{str}(\rho)/V = \min F_{str}\left(\{\rho_i\}, \Gamma_s, \Gamma_D\right),$$  (3.21)

with

$$
\begin{aligned}
\frac{F_{str}\left(\{\rho_i\}, \Gamma_s, \Gamma_D\right)}{T} = & \sum_{i=0}^{f} \rho_i \ln \frac{(f-i)! \, i! \, \rho_i \lambda_T^3}{e} \\
& + f\rho_0 \left(\Gamma_s \ln \Gamma_s + (1-\Gamma_s) \ln(1-\Gamma_s)\right) - \frac{f\rho_0\Gamma_s}{2} \ln \frac{f\rho_0\Gamma_s g_0}{e} \\
& + \rho_D \left(\Gamma_D \ln \Gamma_D + (1-\Gamma_D) \ln(1-\Gamma_D)\right) - \rho_1 \ln \frac{\rho_1 g_0}{e} - \frac{\rho_I}{2} \ln \frac{\rho_I g_0}{e}
\end{aligned}
$$  (3.22)

where the first term is the sum of the free energies of the ideal monomers of all classes with *i=0,1,2* corresponding to the sol, dangling and backbone monomers, respectively, and *i>2* to the junction ones (with due regard for their symmetry indices). The second and third terms in (3.22) are the free energy of choice of the reacted sol groups and that of formation of bonds between them, respectively. The next term is the free energy of choice of the reacted *D*-groups. The last two terms are the free energies of formation of *D-D*[+] (yellow) and *I-I* (blue) bonds, respectively, and the minimum in (3.21) is to be found with due regard for the definitions (3.1), (3.2) and condition of fixed total density of all monomers

$$\rho = \rho_g + \rho_s = \sum_{i=0}^{f} \rho_i$$  (3.23)

the sol monomer density being designated as $\rho_0$. To find the desired conditional minimum of the function (3.21) we find, following Lagrange, the unconditional minimum of the function

$$\hat{F}_{str}(\{\rho_i\}, \Gamma_s, \rho_D, \mu, \nu, \lambda)/T = \sum_{i=0}^{f} \rho_i \ln\left[(f-i)! \, i! \, \rho_i \lambda_T^3 / e\right]$$
$$+ f\rho_0(\Gamma_s \ln \Gamma_s + (1-\Gamma_s)\ln(1-\Gamma_s)) - (f\rho_0\Gamma_s/2)\ln(f\rho_0\Gamma_s g_0/e)$$
$$+ \mu\left(\rho - \sum_{i=1}^{f}\rho_i\right) + \lambda\left(\rho_I - \sum_{i=2}^{f} i\rho_i\right) + \nu\left(\rho_D - \sum_{i=1}^{f}(f-i)\rho_i\right)$$
$$- \rho_1 \ln\frac{\rho_1 g_0}{e} - \frac{\rho_I}{2}\ln\frac{\rho_I g_0}{e} + \rho_1\ln\rho_1 + (\rho_D-\rho_1)\ln(\rho_D-\rho_1) - \rho_D\ln\rho_D$$

(3.24)

where $\mu, \nu, \lambda$ are the Lagrange multipliers to be determined by substituting the extremal values of the structural parameters $\{\rho_i\}, \Gamma_s, \rho_D$ into auxiliary conditions (3.2), (3.23), the condition (3.1) being used already to replace the parameter $\Gamma_D$ in the function (3.24) by $\rho_D$.

As usually, to find a minimum (generally, extremum) of the function $\hat{F}_{str}(\{\rho_i\}, \Gamma_s, \rho_D)$ one should take its derivatives with respect to all independent variables and set them equal to zero. The resulting simultaneous extremal equations determine the equilibrium value of all the relevant structural variables of weak gels within the Flory model. Introducing parameters $\Psi = \exp\lambda$, $z = \lambda_T^{-3}\exp\mu$, $\Phi = \exp\nu$ and new variables reduced by the factor $g_0$ and marked with a wave (e.g., $\tilde{x} = g_0 x$), we can write these equations as follows (the derivatives they are obtained from being also indicated):

$$\partial\hat{F}_{str}/\partial\rho_{i>1} = 0 \quad \rightarrow \quad \tilde{\rho}_{i>1} = \tilde{z}\,\Phi^{f-i}\Psi^i / [i!(f-i)!], \tag{3.25}$$

$$\partial\hat{F}_{str}/\partial\rho_D = 0 \quad \rightarrow \quad \Phi = \rho_D/(\rho_D-\rho_1) \rightarrow \quad \Gamma_D = \rho_1/\rho_D = (\Phi-1)/\Phi \tag{3.26}$$

$$\partial\hat{F}_{str}/\partial\rho_1 = 0 \quad \rightarrow \quad \tilde{\rho}_1\Gamma_D = \tilde{z}\,\tilde{\rho}_1(1-\Gamma_D)\Phi^{f-1}/(f-1)! \tag{3.27}$$

Two important relationships we use below follow from (3.26) and (3.27):

$$\Gamma_D = \tilde{z}\,\Phi^{f-2}/(f-1)! \tag{3.28}$$

$$W(\Phi) = \Phi - 1 - \tilde{z}\,\Phi^{f-1}/(f-1)! = 0 \tag{3.29}$$

Now, for the density $\rho_I$ of all blue (*I-I*) bonds there is the extremal equation

$$\partial\hat{F}_{str}/\partial\rho_I = 0 \quad \rightarrow \quad \Psi^2 = \tilde{\rho}_I \tag{3.30}$$

and an equation stemming just from the definition of the density $\rho_I$:

$$\tilde{\rho}_I = \sum_{i=2}^{f} i\tilde{\rho}_i = \tilde{z}\Psi\left[(\Phi+\Psi)^{f-1} - \Phi^{f-1}\right]/(f-1)! \tag{3.31}$$

Excluding $\tilde{\rho}_I$ from (3.30) and (3.31), we get

$$\Psi = \tilde{z}\left[(\Phi+\Psi)^{f-1} - \Phi^{f-1}\right]/(f-1)! \tag{3.32}$$

Rewriting eq 3.32 with use of (3.29) in the form

$$W(\Psi+\Phi) = W(\Phi) = 0, \tag{3.33}$$

we see that the quantities $\Phi$ and $\Phi+\Psi$ are just two positive roots of equation (3.29), $\Phi$ being the least one, thus determining both $\Phi$ and $\Phi+\Psi$ as some functions of. $\tilde{z}$ . As we show below, this result is just rearrangement of the Flory rule concerning total conversion of the gel phase and that of sol-fraction.

To check this assertion, we introduce the density $\rho_b$ of finite (yellow) branches hanging from the blue trails:

$$\tilde{\rho}_b = \sum_{i=2}^{f} (f-i)\tilde{\rho}_i = \tilde{z}\Phi\frac{\left[(\Phi+\Psi)^{f-1} - \Phi^{f-1} - (f-1)\Psi\Phi^{f-2}\right]}{(f-1)!} = \Phi\Psi\left[1-(f-1)\Gamma_D\right] \tag{3.34}$$

where we used eqs 3.28, 3.32 to get the last of eqs 3.34 that, jointly with definitions

$$\rho_D = \rho_1/\Gamma_D = \rho_b + (f-1)\rho_1 \tag{3.35}$$

gives the expressions for the densities $\rho_I$ and $\rho_D$ of the dangling monomers and all $D$-groups:

$$\tilde{\rho}_1 = (\Phi-1)\Psi = \tilde{z}\,\Psi\,\Phi^{f-1}/(f-1)!, \tag{3.36}$$

$$\tilde{\rho}_D = \Phi\Psi . \tag{3.37}$$

Summing up the densities of all groups belonging to gel-fraction and dividing it by monomer functionality $f$, we get the total density of the monomers belonging to gel-fraction:

$$\tilde{\rho}_g = \left(\tilde{\rho}_I + \tilde{\rho}_b + f\tilde{\rho}_1\right)/f = \tilde{z}\left[(\Phi+\Psi)^f - \Phi^f\right]/f!, \tag{3.38}$$

and that of the unreacted groups belonging to gel-fraction (from (3.26), (3.32) and (3.37)):

$$\tilde{\rho}_{un}^{g} = \tilde{\rho}_{D}\left(1-\Gamma_{D}\right) = \Psi = \tilde{z}\left[\left(\Phi+\Psi\right)^{f-1} - \Phi^{f-1}\right]/(f-1)!. \tag{3.39}$$

To complete, we still are to consider the extremal equations describing equilibrium within the sol-fraction:

$$\partial\hat{F}_{str}/\partial\Gamma_{s} = 0 \quad \rightarrow \quad \Gamma_{s}/\left(1-\Gamma_{s}\right)^{2} = f\,\tilde{\rho}_{0}, \tag{3.40}$$

$$\partial\hat{F}_{str}/\partial\rho_{o} = 0 \quad \rightarrow \quad \tilde{\rho}_{0} = \tilde{z}\left(1-\Gamma_{s}\right)^{-f}/f! \tag{3.41}$$

Excluding from eqs 3.40, 3.41 the sol monomer density, we get the equation (somewhat rearranged eq 3.12c) that determines sol conversion as function of $\tilde{z}$.

$$\tilde{z}\left(1-\Gamma_{s}\right)^{2-f}/(f-1)! = \Gamma_{s} \quad \text{or} \quad \left(1-\Gamma_{s}\right)^{-1} - 1 = \tilde{z}\left[\left(1-\Gamma_{s}\right)^{-1}\right]^{f-1}/(f-1)! \tag{3.42}$$

Comparing equations (3.28), (3.32), (3.40), (3.41), we arrive at the equations

$$\Phi = \left(1-\Gamma_{s}\right)^{-1}, \tag{3.43a}$$

$$\Gamma_{s} = \left(\Phi-1\right)/\Phi = \Gamma_{D} \tag{3.43b}$$

$$\tilde{\rho}_{0} = \tilde{z}\,\Phi^{f}/f! \tag{3.43c}$$

To complete our derivation, we calculate the total monomer density and conversion in the gel phase:

$$\tilde{\rho} = \tilde{\rho}_{g} + \tilde{\rho}_{s} = \tilde{z}\left(\Phi+\Psi\right)^{f}/f! \tag{3.44}$$

$$1-\Gamma = \left(\tilde{\rho}_{un}^{s} + \tilde{\rho}_{un}^{g}\right)/\left(f\,\tilde{\rho}\right) = \left(\tilde{\rho}_{s}\left(1-\Gamma_{s}\right) + \tilde{\rho}_{un}^{g}\right)/\left(f\,\tilde{\rho}\right) = \left(\Phi+\Psi\right)^{-1} \tag{3.45}$$

It follows from eqs 3.29, 3.33, 3.44, 3.45, that the total conversion still obeys the MAL with respect to the total monomer density in gel phase:

$$\Gamma/\left(1-\Gamma\right)^{2} = f\,\tilde{\rho},$$

and is related to sol-fraction (or dangling monomers) conversion by the Flory rule

$$\Gamma_{s}\left(1-\Gamma_{s}\right)^{f-2} = \Gamma\left(1-\Gamma\right)^{f-2} = \tilde{z}/(f-1)! \tag{3.46}$$

Besides, the parameter $\Psi$, that characterizes the distribution of the IC monomers as consistent with (3.25), (3.36), (3.37), can be expressed in terms of both conversions as follows:

$$\Psi = \left(1-\Gamma\right)^{-1} - \left(1-\Gamma_{s}\right)^{-1} \quad \rightarrow \quad \begin{cases} \Psi = 0, & \Gamma < \Gamma_{c} = (f-1)^{-1} \\ \Psi > 0, & \Gamma > \Gamma_{c} \end{cases} \tag{3.47}$$

It follows from (3.47) that the order parameter $\Psi$ monotonously increases from zero value when the total conversion and, thus, the total monomer density of the gel phase increase.

At last, substituting all the structural characteristics into (3.24) one can check that almost all terms cancel each other and the final expression for the weak gel free energy takes the form:

$$F_{str}(\rho)/VT = \rho(\mu/T) - \rho + f\rho\Gamma/2 \qquad (3.48)$$

where the chemical potential $\mu = T\ln z$ defined, as consistent with (3.44), (3.45), as follows:

$$\mu = T\left[\ln(\rho\, f!) + f\ln(1-\Gamma)\right] \qquad (3.49)$$

Substituting (3.49) into (3.48), we finally return to expression (2.40) we derived in Section 2 via the simpler (unstructured) version of the density functional description:

$$F_{str}(\rho, T) = V\, F_{id}(\rho, T) + T\, f\, N\left[(\Gamma/2) + \ln(1-\Gamma)\right]$$

Summarizing, we demonstrated in this section that the broken monomer identity concept, as visualized via the coloring procedure described above, can be made fully consistent with both commonly known classic models of gelation assuming that within any however large but finite window the IC cluster structure is tree-like. Moreover, it is our approach which gives a natural solution of the old problem which model describes the thermoreversible (weak) gelation better: evidently, it would be that model whose free energy is lower. Now, as we just have shown, within the Stockmayer model some relevant structural variables (the junction monomer densities $\rho_i$, $i>2$) are disregarded and arbitrarily equaled to zero, whereas within the Flory model these variables take certain finite values as found via minimization of the free energy. We conclude that the Flory model has the lower free energy and is, therefore, more adequate than ST model (the corresponding plots are presented in section 5).

At this point we encounter an entirely new problem. Does Flory description, indeed, take into account *all the relevant* structural parameters of the IC? Or are there some new structural parameters, whose account would enable us *still to diminish* the equilibrium free energy as compared to the Flory one? This problem was first addressed by Erukhimovich[25] and solved, in a consistent post-gel meanfield approximation, in ref 29. In the next Section we present the more detailed consideration of this problem via the BMI concept.

## 4. The BMI concept and the structured density functional description of the IC including mesoscopic cycles.

*4.1. The density functional description of the mesoscopically cycled IC.* The starting point for our analysis in this section is the seemingly trivial fact that an infinite cluster without cycles, unlike a finite tree-like cluster, cannot be embedded into 3D space. In the more quantitative terms, it follows from the fact that the perturbation corrections to the mean field Flory approximation, i.e. the contributions into the structural free energy due to complicated cyclic fragments, are characterized[24,25] by the cyclization parameter

$$\kappa = \left( f \rho a^3 \right)^{-1} \left( \Gamma / \Gamma_c \right)^3 / \left( 1 - \Gamma / \Gamma_c \right)^{3/2} \qquad (4.1)$$

This parameter diverges at the classic sol-gel threshold and, therefore, the cyclization effects become dominant in a close vicinity of the threshold and can be expected to be dominating in the gel-fraction. Putting it in other words, the structure of the IC still can be represented as Cayley tree on fairly large scales. However, the IC includes not only the "bare" vertices of the $n$-th order, $1 \le n \le f$, but also the peculiar "effective" vertices of arbitrary order and complexity having form of 1-irreducible blocks depicted in Figure 1d.

Such a description of the IC structure[24,25] is somewhat similar to the droplet model of the infinite cluster[32,33]. Unlike the latter, however, we focus our attention, instead of the self-similar structure of the 1-irreducible blocks, on the possibility to distinguish between the "internal" bonds, from which these blocks are constructed, and "external" ones, that join blocks into an effective Cayley tree. To make such a distinction we adopt a new sort of the coloring procedure as follows.

First, we choose a finite window and color all the bonds A-A within that window yellow, red and blue as described in the previous section but the following distinction. If a bond between two groups belongs to, at least, one closed trail of bonds confined to the window, we color it green and call C-C bond to indicate that it belongs to cyclic fragments of the IC (see Figure 4a). As the size of the window is increased, some of the initially blue bonds turn out, at fairly large scales $L$, to belong to a closed path and should be recolored green. Thus, as $L$ increases, the fraction of green groups grows and reaches a certain limiting value in the limit $L \to \infty$ when all the "internal" bonds are colored green and the "external" bonds of the IC are colored blue, yellow or red as explained in the previous section.

The coloration procedure just described does not alter the statistical weights and symmetry indices of the various diagram realizations of the structure of the infinite cluster. But it is very important for implementing an approximate "mean-field" summation of the complex cycled fragment contributions to the structural term $F_{str}$ that in many respects is similar to calculation of the high-order diagram contributions to the Gell-Mann-Low function.[34]

Indeed, as shown in ref 25, when the first of conditions (2.18) holds, the typical blocks determining the IC structure are the "bare" vertices (Figure 1a) and very complicated blocks (Figure 1d gives a rather rough idea about such blocks structuring and composing into larger fragments). Therewith, the contribution of comparatively simple cyclic blocks (like those shown in Figures 1b,c) can be neglected. But if we consider only a part of a large complicated cluster confined to a window large as compared to the bond length $l$ but small in comparison to the size of the whole block (see Figure 4a), this part can appear as a Cayley tree. The true topology of such a quasi-Cayley tree (its inclusion in the system of closed trails of bonds) is intimated, as shown in Figure 4c, by a local coloring which indicates the number $i$ of green groups belonging to each monomer in an assigned realization of the infinite cluster. Note that $i$ runs through all integer values in the range $0 \le i \le f$ except for $i=1$ (a monomer cannot be included into a closed path via only one group).

The next step is to assume that an adequate evaluation of the total structural free energy $F_{str}$ for the weak gels could be achieved just via calculation of those contributions that correspond to the formation and all possible recombinations of the bonds within windows of not too large size (less than a typical size of the complicated cycled fragment) followed by summation of these contributions. Of course, the effects caused by the correlation of the structure of neighboring windows (i.e., the hierarchical structure of the blocks) are ignored within such a "mean-field" description of the infinite cluster, but, in return, we can take into account the effects which are caused by the difference between "internal" and "external" bonds described above.

Indeed, the fact itself that a trail of bonds is closed (even somewhere far beyond the window under consideration) alters the combinatorial weight of this trail (the number of different ways to form it). It is this change in combinatorial behavior which is taken into account, at the level of the coarse-grained density functional description, by considering vertices (monomers) of different colors and corresponding altering their symmetry index, as seen when comparing the coloration of the same quasi-Cayley tree corresponding to the Flory and mesoscopic cyclization (MC) models.

More precisely, there are three types of chemical bonds (green $C$-$C$, blue $I$-$I$ and yellow $D$-$D^+$) that belong to the IC. Accordingly, there are only the following topological types of the IC monomers (see Figure 3): 1) the dangling monomers $D^+D_{f-1}$ that form a fringe hanging from the IC backbone (all the bonds adjacent to such monomers are yellow), 2) the monomers $C_mD_{f-m}$ with $2 \leq m \leq f$ that belong to the cycled fragments of the IC ($m$ bonds adjacent to such monomers are green and the rest of them are yellow), 3) the monomers $I_mD_{f-m}$ with $2 \leq m \leq f$ that belong to the tree-like fragments of the IC ($m$ bonds adjacent to such monomers are blue and the rest of them are yellow), and 4) the junction monomers $C_nI_mD_{f-m-n}$ (with $n \geq 2$, $m \geq 1$, $n+m \leq f$) that implant the mesoscopic cycles into the infinite cluster.

Thus, within the BMI concept-based structured density functional description, first we determine a state of the system quantitatively by fixing the average numbers per unit volume of the dangling monomers $\rho_1$, those with $n$ green and $m$ blue bonds $\rho_{n,m}$ and sol monomers $\rho_0$.

Next, we calculate the combinatorial contributions to the free energy for such a fixed structured densities distribution, and, finally, determine the thermodynamically equilibrium values of these densities requiring that the total free energy should be minimal with respect to the latter. If such a minimization would result in zero total density $\rho_C$ of the bonds belonging to the closed trails (as it happened to $\rho_2$ within the ST model), we would prove that no extra structural parameters improve the density functional description provided by the Flory model. However, it is not the case and the thermodynamically equilibrium value of $\rho_C$ turns out to be finite.[25,29] This fact changes the conventional picture of sol-gel transition drastically. In the rest of this section we present a new derivation of the structural free energy of weak gels including also description of the structure of the mesoscopically cycled weak gels.

As evident from the preceding discussion, the desired structural free energy of weak gels, with due regard for MC effects, takes the following form:

$$F_{str}(\rho)/V = \min F_{str}(\rho, \rho_C), \tag{4.2}$$

where we introduced a specific structural free energy $F_{str}(\rho, \rho_C)$ for a partially equilibrium state of weak gel with a fixed value $\rho_C$ of the green groups density. It is defined as follows:

$$F_{str}(\rho, \rho_C) = (\rho_C/2)\ln(e/(\rho_C g_0)) + \overline{F}_{str}(\rho, \rho_C) \tag{4.3}$$

The first term in (4.3) is the specific free energy of the green bonds formation and

$$\begin{aligned}
F_{str}(\rho, \rho_C) = \min\Bigg\{ &\rho_0 \ln \frac{f!\rho_0\lambda_T^3}{e} + \rho_1 \ln \frac{(f-1)!\rho_1\lambda_T^3}{e} + \sum_{ij}\rho_{ij} \ln \frac{(f-i-j)!i!j!\rho_{ij}\lambda_T^3}{e} \\
&+ f\rho_0\left(\Gamma_s \ln \Gamma_s + (1-\Gamma_s)\ln(1-\Gamma_s)\right) + \rho_D\left(\Gamma_D \ln \Gamma_D + (1-\Gamma_D)\ln(1-\Gamma_D)\right) \\
&- \rho_1 \ln \frac{\rho_1 g_0}{e} - \frac{f\rho_0\Gamma_s}{2}\ln\frac{f\rho_0\Gamma_s g_0}{e} - \frac{\rho_I}{2}\ln\frac{\rho_I g_0}{e} \Bigg\}
\end{aligned} \tag{4.4}$$

where the first two terms (4.4) are the free energies of the ideal sol and dangling monomers, respectively, the third term is the sum of those for all the backbone and junction monomers forming both the tree-like and cycled fragments of the IC (with due regard for their symmetry indices). The fourth and fifth terms in (4.4) are the free energy of choice of the reacted *S*- and *D*-groups. The last three terms (4.4) are the free energies of formation of yellow, red and blue bonds, respectively. The summation in (4.4) and henceforth is implied over all pairs $(i,j)$ with $i=0,2,..f$, $j=0,1..f$ (the number $i$ being always corresponded to the number of green bonds adjacent to the monomer) but the pairs $(0,0)$ and $(0,1)$ equivalent to the accounted separately sol and dangling monomers, respectively. The minimum in (4.4) is to be found with due regard for condition of fixing the total density of all monomers $\rho$ and that of green groups $\rho_C$, definitions (3.1)-(3.2) and those of the densities of all *I*-groups (blue) and *C*-groups (green):

$$\rho_{D^+} = \Gamma_D \rho_D = \rho_1, \qquad \rho_D = (f-1)\rho_1 + \sum_{ij} (f-i-j)\rho_{ij}, \qquad (4.5a)$$

$$\rho_I = \sum_{ij} j \rho_{ij}, \qquad \rho_C = \sum_{ij} i \rho_{ij}, \qquad \rho = \rho_S + \rho_1 + \sum_{ij} \rho_{ij} \qquad (4.5b)$$

At last, the minimum in (4.3) is to be sought within the interval $0 \le \rho_C \le f\rho$.

### 4.2. Finite and infinite edge length approximations to describe the MC effects.

Actually, the MC effects have been considered in two different approximations.[25,29] To describe them, we distinguish monomers of sort $(2,n)$, $n=0,1..f$, and those of sort $(i,n)$, $2<i\leq f$, $0\leq n\leq f-i$, called further the backbone and junction green monomers (BGM and JGM) respectively. In the original treatment[25] only the tree-like and backbone green monomers were taken into account, for simplicity, whereas in the later study[29] we have allowed for all sorts of the permissible green monomers.

The difference between the approximations is most transparently visualized by the notion of the green edges, i.e. the trails of BGM between the nearest JGM, whose length is defined as the average number of the BGM per green edge. Noticing that the number of green edges is just one half of the number of green groups belonging to junction green monomers we call further the junction green FG (JGFG), one can write

$$l_{\text{edge}} = \rho_{\text{BGM}}/\rho_{\text{green edge}} = 2\rho_{\text{BGM}}/\rho_{\text{JGFG}} = 2\rho_{\text{BGM}}/(\rho_C - 2\rho_{\text{BGM}}), \qquad (4.6)$$

where $2\rho_{BGM}$ and $\rho_{JGFG}$ are densities of the green groups belonging to BGM and JGM, respectively. Thus, the edge length is infinity if $\rho_{JGFG}$ is assumed to equal zero, and it is finite if $\rho_{JGFG}>0$. So, we will refer to approximations of refs 25 and 29 as the infinite and finite edge approximations (IEA and FEA), respectively, and consider both in parallel way.

*4.3. Calculation of the structural characteristics and free energy of mesocycled gel.* To calculate the function (4.4), we introduce the Lagrange multipliers $\mu,\nu,\lambda,\gamma$ associated with the total monomer density $\rho$ and the densities $\rho_D$ of active yellow (fringe) and $\rho_I$, $\rho_C$ of all blue and green (belonging to tree-like and cycled fragments of the IC cluster backbone) FG, respectively and find the unconditional minimum of the function

$$
\begin{aligned}
\hat{F}_{str}\left(\rho,\rho_0,\rho_1\{\rho_{ij}\}\Gamma_s,\rho_D\right)/T &= \sum_{ij}\rho_{ij}\ln\frac{(f-i-j)!i!j!\rho_{ij}\lambda_T^3}{e} + \rho_1\ln\frac{(f-1)\rho_1\lambda_T^3}{e} \\
&+ \rho_0\ln\frac{f!\rho_0\lambda_T^3}{e} + f\rho_0\left(\Gamma_s\ln\Gamma_s+(1-\Gamma_s)\ln(1-\Gamma_s)\right) + \rho_1\ln\rho_1 + (\rho_D-\rho_1)\ln(\rho_D-\rho_1) \\
&- \rho_D\ln\rho_D - -\frac{f\rho_0\Gamma_s}{2}\ln\frac{f\rho_0\Gamma_s g_0}{e} - \rho_1\ln\frac{\rho_1 g_0}{e} - \frac{\rho_I}{2}\ln\frac{\rho_I g_0}{e} - \lambda\left(\rho_I - \sum_{ij}j\rho_{ij}\right) \\
&- \mu\left(\rho-\rho_0-\rho_1-\sum_{ij}\rho_{ij}\right) - \gamma\left(\rho_C-\sum_{ij}i\rho_{ij}\right) - \nu\left(\rho_D-(f-1)\rho_1-\sum_{ij}(f-i-j)\rho_{ij}\right)
\end{aligned}
\tag{4.7}
$$

To determine $\mu,\nu,\lambda,\gamma$ we substitute the extremal values of the structural parameters $\rho_0,\rho_I$, $\{\rho_{ij}\}\Gamma_s,\rho_D,\Gamma_D$ into auxiliary conditions (4.5), the first of which being used already to replace the parameter $\Gamma_D$ in the function (4.4) by $\rho_D$. The resulting from (4.7) simultaneous extremal equations determine the equilibrium values of all the relevant structural variables for our new model of mesocycled weak gels.

Our strategy in performing the desired minimization of the free energy is as follows: i) we introduce, by analogy with consideration of the Flory model in the preceding section, the parameters $z=\lambda_T^{-3}\exp\mu$, $\Phi=\exp\nu$, $\Psi=\exp\lambda$ and a new parameter $\Theta=\exp\gamma$, ii) express all the waved (reduced by the factor $g_0$) structural variables in terms of these parameters via the extremal equations for the function (4.7), iii) find, using the constraints (4.5), a convenient one-parametric representation of all the variables and parameters as functions of $\tilde{\rho}$ and $\tilde{\rho}_C$, iv) minimize the partially equilibrium free energy (4.3) with respect to $\tilde{\rho}_C$ and find the structural free energy and some other weak gel characteristics as a parametric function of $\tilde{\rho}$.

*4.3.1. Calculation of the partially equilibrium free energy* (4.3) *of mesocycled gel.* To begin with, let us write the corresponding simultaneous extremal equations (the derivatives they are obtained from are also indicated):

$$\partial \hat{F}_{str} / \partial \rho_{ij} = 0 \quad \rightarrow \quad \tilde{\rho}_{ij} = \tilde{z} \Phi^{f-i-j} \Psi^j \Theta^i \Big/ \big[ i! \, j! (f - i - j)! \big] \tag{4.8}$$

$$\partial \hat{F}_{str} / \partial \rho_D = 0 \quad \rightarrow \quad \Phi = \rho_D / (\rho_D - \rho_1) \rightarrow \Gamma_D = \rho_1 / \rho_D = (\Phi - 1) / \Phi \tag{4.9}$$

$$\partial \hat{F}_{str} / \partial \rho_1 = 0 \quad \rightarrow \quad \tilde{\rho}_1 \Gamma_D = \tilde{z} \tilde{\rho}_1 (1 - \Gamma_D) \Phi^{f-1} / (f - 1)! \tag{4.10}$$

Two important relationships we use below follow from (4.9) and (4.10):

$$\Gamma_D = \tilde{z} \Phi^{f-2} / (f - 1)! \tag{4.11}$$

$$\tilde{z} / (f - 1)! = (\Phi - 1) / \Phi^{f-1} = \Gamma_D (1 - \Gamma_D)^{f-2} \tag{4.12}$$

The next group of equations is related to the density $\rho_I$ of all blue bonds:

$$\partial \hat{F}_{str} / \partial \rho_I = 0 \quad \rightarrow \quad \tilde{\rho}_I = \Psi^2 \tag{4.13}$$

and the monomer density and conversion within the sol-fraction:

$$\partial \hat{F}_{str} / \partial \Gamma_s = 0 \quad \rightarrow \quad \Gamma_s / (1 - \Gamma_s)^2 = f \tilde{\rho}_0, \tag{4.14}$$

$$\partial \hat{F}_{str} / \partial \rho_0 = 0 \quad \rightarrow \quad \tilde{\rho}_0 = \tilde{z} (1 - \Gamma_s)^{-f} / f! \tag{4.15}$$

Excluding the sol monomer density $\rho_0$ from the latter yields sol conversion as function of $\tilde{z}$:

$$\tilde{z} / (f - 1)! = \Gamma_s (1 - \Gamma_s)^{f-2} \tag{4.16}$$

Comparing eqs 4.9, 4.12 and 4.16 we get

$$\Gamma_s = \Gamma_D = (\Phi - 1) / \Phi \tag{4.17}$$

$$f \tilde{\rho}_0 = \tilde{z} \Phi^f / (f - 1)! = \Phi (\Phi - 1) = \Gamma_s / (1 - \Gamma_s)^2 \tag{4.18}$$

The densities $\rho_I$ and $\rho_C$ are also determined from the definitions (4.5b), the form of the corresponding expressions being determined by choice of approximation we use: in the IEA both $\rho_I$ and $\rho_C$ include the terms of the zeroth and second power in $\Theta$ only, whereas in the FEA they contain all the terms up to the $f$-th power in $\Theta$ but the linear one. Thus, we get

$$\tilde{\rho}_I = \sum_{ij} j\tilde{\rho}_{ij} = \begin{cases} \tilde{z}\,\Psi \left[ \dfrac{(\Phi+\Psi)^{f-1} - \Phi^{f-1}}{(f-1)!} + \dfrac{\Theta^2}{2}\dfrac{(\Phi+\Psi)^{f-3}}{(f-3)!} \right] & IEA \\[3mm] \tilde{z}\,\Psi\,\dfrac{(\Phi+\Psi+\Theta)^{f-1} - (f-1)\Theta(\Phi+\Psi)^{f-2} - \Phi^{f-1}}{(f-1)!} & FEA \end{cases} \tag{4.19}$$

$$\tilde{\rho}_C = \sum_{ij} i\tilde{\rho}_{ij} = \begin{cases} \tilde{z}\,\Theta^2(\Phi+\Psi)^{f-2}/(f-2)! & IEA \\[2mm] \tilde{z}\,\Theta\left[(\Phi+\Psi+\Theta)^{f-1} - (\Phi+\Psi)^{f-1}\right]/(f-1)! & FEA \end{cases} \tag{4.20}$$

In the post-gel region, where $\tilde{\rho}_I$ and $\Psi$ differ from zero, one can exclude $\tilde{\rho}_I$ from eqs 4.13, 4.19 which leads to equalities

$$\Psi = \begin{cases} \tilde{z}\left[ \dfrac{(\Phi+\Psi)^{f-1} - \Phi^{f-1}}{(f-1)!} + \dfrac{\Theta^2}{2}\dfrac{(\Phi+\Psi)^{f-3}}{(f-3)!} \right] & IEA \\[3mm] \tilde{z}\left[(\Phi+\Psi+\Theta)^{f-1} - (f-1)\Theta(\Phi+\Psi)^{f-2} - \Phi^{f-1}\right]/(f-1)! & FEA \end{cases} \tag{4.21}$$

Next, we introduce a new structural variable, the density $\rho_b$ of finite (yellow) branches hanging from the blue and green trails, whose definition follows from (4.4a):

$$\rho_b = \rho_D\left(1 - (f-1)\Gamma_D\right) = \sum_{ij}(f-i-j)\rho_{ij} \tag{4.22}$$

Using formulas (4.8), (4.11) and (4.21), the density $\rho_b$ is calculated as follows:

$$\begin{aligned} \tilde{\rho}_b &= \sum_{ij}(f-i-j)\tilde{\rho}_{ij} = \tilde{z}\Phi\,\frac{(\Phi+\Psi+\Theta)^{f-1} - (f-1)\Theta(\Phi+\Psi)^{f-2} - (f-1)\Phi^{f-2}\Psi - \Phi^{f-1}}{(f-1)!} \\ &= \Phi\Psi\left[1 - (f-1)\Gamma_D\right] \end{aligned} \tag{4.23}$$

Substituting this result back to (4.22) and (4.5a) gives the expressions for the densities $\rho_I$ and $\rho_D$ of the dangling monomers and all $D$-groups, respectively:

$$\tilde{\rho}_D = \Phi\Psi \tag{4.24}$$

$$\tilde{\rho}_1 = \tilde{\rho}_{D^+} = (\Phi-1)\Psi = \tilde{z}\Psi\Phi^{f-1}/(f-1)! \tag{4.25}$$

Now, using (4.8), (4.18), (4.25), we can express in terms of the parameters $z, \Phi, \Psi, \Theta$ the total

monomer density:

$$\tilde{\rho} = \tilde{\rho}_0 + \tilde{\rho}_1 + \sum_{ij} \rho_{ij} = \tilde{z} A_f(\delta) \frac{(\Phi + \Psi)^f}{f!}, \quad A_f(\delta) = \begin{cases} 1 + f(f-1)\delta^2/2 & \text{IEA} \\ (1+\delta)^f - f\delta & \text{FEA} \end{cases} \quad (4.26)$$

where we introduced a parameter $\delta = \Theta/(\Phi + \Psi)$ describing a state of mesoscopic cycling, the density of those FG (red, yellow and blue) that belong to tree-like fragments of both the IC and sol fraction (and, therefore, are not included into mesoscopic cycles):

$$\tilde{\rho}_{tot}^{tree} = f\tilde{\rho}_0 + \tilde{\rho}_{D^+} + \tilde{\rho}_D + \tilde{\rho}_I = (\Phi + \Psi)(\Phi + \Psi - 1), \quad (4.27a)$$

and the conversion of these tree-like fragments called further the external conversion:

$$\Gamma_e = 1 - \left[(1-\Gamma_s)\tilde{\rho}_0 + (1-\Gamma_D)\tilde{\rho}_D\right]/(\tilde{\rho}_{tot}^{tree}) = 1 - (\Phi + \Psi)^{-1} \quad (4.27b)$$

As seen from eqs 4.27, the external conversion is related to the density $\tilde{\rho}_{tot}^{tree}$ via the MAL

$$\tilde{\rho}_{tot}^{tree} = f\tilde{\rho} - \tilde{\rho}_C = f\tilde{\rho}(1-\Gamma_i) = \Gamma_e/(1-\Gamma_e)^2 \quad (4.28)$$

where $\Gamma_i$ is the fraction of groups belonging to mesoscopic trails called internal conversion.

We also easily arrive at expression for the density of unreacted groups in the sol fraction:

$$\tilde{\rho}_s^{un} = (1-\Gamma_s)f\tilde{\rho}_0 = \Phi - 1, \quad (4.29a)$$

that for all unreacted groups both in sol and gel fractions:

$$\tilde{\rho}_{tot}^{un} = (1-\Gamma_e)\tilde{\rho}_{tot}^{tree} = \Phi + \Psi - 1, \quad (4.29b)$$

and thus at that for unreacted groups belonging to the gel fraction only:

$$\tilde{\rho}_g^{un} = \tilde{\rho}_{tot}^{un} - \tilde{\rho}_s^{un} = \Psi. \quad (4.29c)$$

To find a relationship between sol and external conversions, we notice that $\tilde{\rho}_{tot}^{tree}$ may be rewritten, using eqs 4.20, 4.26, as follows:

$$\tilde{\rho}_{tot}^{tree} = f\tilde{\rho} - \tilde{\rho}_C = \tilde{z}(\Phi + \Psi)^f A_{f-1}(\delta)/(f-1)!, \quad (4.30)$$

Now, substituting (4.27), (4.28) into (4.30) and using (4.16) we get the reduced fugacity $\tilde{z}$ and sol conversions $\Gamma_s$ as functions of $\delta$ and external conversion $\Gamma_e$:

$$\tilde{z}/(f-1)! = \Gamma_e(1-\Gamma_e)^{f-2}/A_{f-1}(\delta), \quad (4.31)$$

$$\Gamma_s\big(\tilde{z}(\delta)\big)\big(1 - \Gamma_s\big(\tilde{z}(\delta)\big)\big)^{f-2} = \Gamma_e\big(1 - \Gamma_e\big)^{f-2}\Big/A_{f-1}(\delta) \qquad (4.32)$$

As is seen from eq 4.32, the Flory rule (3.45) holds if $A_f(\delta)=1$ which takes place only if $\delta=0$ and, therefore, no MC occurs. Otherwise a finite value of $\delta$ is determined from the expression for the internal conversion $\Gamma_i$ that follows from eqs 4.20, 4.26:

$$\Gamma_i = \big(f\tilde{\rho} - \tilde{\rho}_{tot}^{tree}\big)\big/\big(f\tilde{\rho}\big) = 1 - \big(A_{f-1}(\delta)/A_f(\delta)\big) = \delta\big[(1+\delta)^{f-1} - 1\big]\big/\big[(1+\delta)^f - f\delta\big]. \qquad (4.33)$$

Calculating the function $F_{str}\big(\rho, \rho_C\big)$ by substituting equilibrium values of quantities appearing in its definition into (4.4) and (4.3), we get

$$F_{str}\big(\rho, \rho_C\big) = F_{id}\big(\rho\big) + T\Big\{f\rho\big[\Gamma/2 + \ln(1 - \Gamma_e)\big] + \big(\rho_C/2\big)\ln\big(\Theta^2/\tilde{\rho}_C\big) - f^{-1}\ln\big(A_f(\delta)\big)\Big\} \qquad (4.34)$$

where the total conversion $\Gamma$ and parameter $\Theta$ are defined as follows:

$$1 - \Gamma = \tilde{\rho}_{tot}^{un}\big/\big(f\tilde{\rho}\big) = \big(1 - \Gamma_e\big)\tilde{\rho}_{tot}^{tree}\big/\big(f\tilde{\rho}\big) = \big(1 - \Gamma_e\big)\big(1 - \Gamma_i\big) \qquad (4.35)$$

$$\Theta = \delta\big(\Phi + \Psi\big) = \delta/\big(1 - \Gamma_e\big) \qquad (4.36)$$

Eqs 4.28, 4.33, 4.35, and 4.37 define $\Gamma_e$, $\Gamma$, $\delta$ and $\Theta$ as some implicit functions of the total reduced monomer density $\tilde{\rho}$ and internal conversion $\Gamma_i$ which enables us to define the specific structural free energy $F_{str}\big(\rho, \rho_C\big)$ as a function of $\Gamma_i$ at a fixed value of $\tilde{\rho}$.

To show explicitly how much is the MC favorable or disadvantageous, we present plots of the MC increment of the free energy $\varphi_{meso} = \big(F_{str}\big(\rho, \rho_C\big) - F_{str}\big(\rho, 0\big)\big)/\big(\rho T\big)$ for $f=3$ and different values of $\tilde{\rho}_{tot} = f\tilde{\rho}$ (see Figure 5).

< Figures 5>

The plots calculated in the IEA are presented in Figure 5a. It is seen that in this approximation the MC (increasing of $\Gamma_i$) leads to increasing of $F_{str}\big(\rho, \rho_C\big)$ (and thus is absolutely disadvantageous) for $\tilde{\rho}_{tot} < \tilde{\rho}_c$. On the contrary, for $\tilde{\rho}_{tot} > \tilde{\rho}_c$ the Flory (tree-like) state of gel-fraction turns out to be absolutely unstable with respect to mesocycling and the equilibrium value of $\Gamma_i$, i.e. the location of the minimum of $F_{str}\big(\rho, \rho_C\big)$, increases monotonously with $\tilde{\rho}_{tot}$ from the value $\Gamma_i^{eq}(\tilde{\rho}_c) = 0$. Such a behavior is characteristic of a second-order phase transition with the parameter $\delta$ playing the role of an order parameter[25].

The situation is different in the FEA with the junction green monomers taken into account. In this approximation there is an interval $\tilde{\rho}_{min} < \tilde{\rho}_{tot} < \tilde{\rho}_c$ where both the Flory (tree-like) and mesocycled states may exist (at least as metastable ones). As seen in Figure 5b, in a vicinity of the classical SGT threshold $\tilde{\rho}_{tot} = \tilde{\rho}_c$ the increments $\varphi$ reveal a comparatively deep negative minimum in whose vicinity the mesocycled states are more thermodynamically advantageous than the Flory (tree-like) state (for $\tilde{\rho}_{tot} < \tilde{\rho}_c$ the latter corresponds to sol phase). The barrier separating the sol and mesocycled gel phase is here low and it disappears in the classical post-gel region $\tilde{\rho}_{tot} \geq \tilde{\rho}_c$. Thus, the latter turns out to be always mesocycled. The further evolution of the increment behavior is shown in Figure 5c. The minimal value of the structural free energy corresponding to the mesocycled gel increases with decrease of $\tilde{\rho}$ until it disappears at $\tilde{\rho}_{tot} = \tilde{\rho}_{min} \approx 0.905\tilde{\rho}_c$ (curve 2 in Figure 5c). In the region $\tilde{\rho}_{tot} < \tilde{\rho}_{min}$ no minimum of the function $\varphi(\Gamma_i)$ exists, therefore, the mesocycled gel is here absolutely unstable with respect to decomposition into the sol phase. Such a behavior is characteristic of the first-order phase transition occurring at $\tilde{\rho}_{tot} = \tilde{\rho}_{SGT} \approx 0.915\tilde{\rho}_c$ (curve 3 in Figure 5c).

So, the distinction between the IEA and FEA demonstrated in Figure 5 is a counterpart of that between the ST and Flory descriptions discussed in Section 3: in both cases the order of the sol-gel transition is determined by accounting or neglecting of junction monomers.

*4.3.1. Calculation of the equilibrium free energy* (4.2) *of mesocycled gel.* In the preceding subsection we proved that there exists a favorable state of MC providing a minimal value of the structural free energy. In this subsection we complete our analysis by deriving a one parameter representation of all relevant thermodynamically equilibrium (i.e. corresponding to this minimum) weak gel characteristics.

To this end we append extremal equations (4.9)-(4.10) and (4.13)-(4.15) by one more extremal equation with respect to the green groups density:

$$\partial \hat{F}_{str} / \partial \rho_C = 0 \quad \rightarrow \quad \tilde{\rho}_C = \Theta^2 \tag{4.37}$$

Comparing eqs 4.20 and 4.37 we get a new equality

$$1 = \tilde{z} \, B_f(\delta) \, (\Phi + \Psi)^{f-2} / (f-1), \qquad B_f(\delta) = (A_f(\delta) - A_{f-1}(\delta)) / \delta^2 \tag{4.38}$$

(remind that this equality holds only for mesoscopically cycled gel with $\delta \neq 0$).

On the other hand, it follows from eqs 4.27a, 4.37 that the total density of the FG $A$ reads

$$\tilde{\rho}_{tot} = f \tilde{\rho} = \tilde{\rho}_{tot}^{tree} + \tilde{\rho}_C = (\Phi + \Psi)(\Phi + \Psi - 1) + \Theta^2 , \tag{4.39}$$

which with eq 4.26 gives the equilibrium external conversion as an explicit function of $\delta$:

$$\Gamma_e(\delta) = 1 - (\Phi + \Psi)^{-1} = A_{f-1}(\delta) / B_f(\delta), \tag{4.40}$$

Substituting (4.40) into (4.35), (4.39) we complete construction of the desired one-parametric representation for all the relevant structural variables:

$$\tilde{\rho}_{tot} = \left[ \Gamma_e(\delta) + \delta^2 \right] / (1 - \Gamma_e(\delta))^2 , \tag{4.41}$$

$$\Gamma = 1 - \frac{(1 - \Gamma_s)\tilde{\rho}_0 + (1 - \Gamma_D)\tilde{\rho}_D}{\tilde{\rho}_{tot}} = \frac{(\Phi + \Psi - 1)^2 + \Theta^2}{(\Phi + \Psi)(\Phi + \Psi - 1) + \Theta^2} = \frac{(\Gamma_e(\delta))^2 + \delta^2}{\Gamma_e(\delta) + \delta^2} \tag{4.42}$$

It follows from eq 4.42 that $\Gamma = \Gamma_e$ in case of $\delta = 0$ which reduces (4.41) to the conventional MAL (2.39). Thus, eqs 4.41, 4.42 generalize this law for mesocycled weak gels.

Two more interesting structural characteristics are the fraction of those monomers that have at least two green groups and thus are included into mesoscopic cycles

$$w_{GM} = \frac{\sum_{i \geq 2} \rho_{ij}}{\rho} = \begin{cases} f(f-1)\delta^2 / 2 & \text{IEA} \\ \left[ (1+\delta)^f - f\delta - 1 \right] / \left[ (1+\delta)^f - f\delta \right] & \text{FEA} \end{cases} \tag{4.43}$$

and that of intramolecular bonds forming the gel-fraction (the reduced cyclomatic index[23,25]):

$$r = \lim_{N_g^0 \to \infty} \left( N_g - N_g^0 \right) / \left[ f(N_g^0 + 1)/2 \right] = \Gamma_g - (2/f) \tag{4.44}$$

The numerator of (4.44) is just the excess of the actual number $N_g$ of chemical bonds forming a macromolecule consisting of $N_g^0 + 1$ monomers as compared to the minimal number of the bonds $N_g^0$ necessary to conjunct these monomers into one macromolecule, known to equal the number of independent cycles (cyclomatic rank) for the macromolecule. The gel fraction conversion $\Gamma_g$ appearing in (4.44) is defined as follows:

$$\Gamma_g = 2N_g \big/ \left(f N_g^0\right) = 1 - \tilde{\rho}_g^{un} \big/ \tilde{\rho}_g^{tot} = \left[\Psi\left(\Psi + 2\Phi - 2\right) + \Theta^2\right] \big/ \left[\Psi\left(\Psi + 2\Phi - 1\right) + \Theta^2\right]$$
$$= \frac{\delta^2\left(1 - \Gamma_s(\delta)\right)^2 + \left(\Gamma_e(\delta) - \Gamma_s(\delta)\right)\left[\Gamma_e(\delta) + \Gamma_s(\delta) - 2\Gamma_e(\delta)\Gamma_s(\delta)\right]}{\delta^2\left(1 - \Gamma_s(\delta)\right)^2 + \left(\Gamma_e(\delta) - \Gamma_s(\delta)\right)\left[1 - \Gamma_e(\delta)\Gamma_s(\delta)\right]} \quad (4.45)$$

Finally, substituting (4.37) into (4.34) we one can check that the structural free energy of the mesocycled weak gels is again given by expression (3.48):

$$F_{str}(\rho)/VT = \rho(\mu/T) - \rho + f\rho\Gamma/2 \quad (4.46)$$

where the chemical potential is determined from (4.26) as follows:

$$\mu/T = \ln\left(\tilde{z}\lambda_T^3 / g_0\right) = \ln f! \rho\lambda_T^3 \ + f \ln\left(1 - \Gamma_e(\delta)\right) - \ln A_f(\delta) \quad (4.47)$$

and the concentration dependence of the total and external conversions for the mesocycled weak gels are parametrically determined by eqs 4.40-4.42. Thus, we reduce (4.46) to the form

$$F_{str}(\rho) = VF_{id}(\rho) + f\,N\,F_{mesocycle}(\tilde{\rho}_{tot}),$$
$$F_{mesocycle}(\tilde{\rho}_{tot})/T = \left(\Gamma/2\right) + \ln\left(1 - \Gamma_e(\delta)\right) - f^{-1}\ln A_f(\delta) \quad (4.48)$$

The reduced concentration dependences of the specific cluster contributions (per one FG) into the structural free energy (4.2) are determined by eqs (2.40a), (3.20b) and (4.48) for all the approaches under discussion and plotted in Figure 6.

< Figures 6>

It is seen from Figure 6 that the FEA has a unique feature of detecting the gel mesocycled phase below the classic SGT in the interval $\tilde{\rho}_{min} < \tilde{\rho}_{tot} < \tilde{\rho}_c$ of the reduced group $A$ densities whereas for all other approximations the specific structural free energies are monotonous and the SGT occurs at the same values of $\Gamma$ and $\tilde{\rho}$. In the region of common existence of all four approximations $\tilde{\rho}_{tot} > \tilde{\rho}$ the FE and ST approximations always provide the larger and lowest, respectively, free energy of the gel-fraction than the Flory one. The function $F_{mesocycle}^{finite}(\tilde{\rho}_{tot})$ defining the extremal values of the specific structural free energy (4.43), reveals in this interval a peculiar singularity of the swallow-tail type (see insets in Figure 6). (Such a singularity was found also when considering equilibrium between a solution of simple cycles and the Flory gel phase in a rather different telechelic system.[35])

It is also seen from Figure 6 that even the simplest approximation (IEA), that takes into account the MC effects, provides a decrease in the specific structural free energy as compared to that of Flory (increment) of the same order of magnitude as the latter does in comparison to the Stockmayer treatment. Notice at last, that the absolute values of the increments increase when functionality $f$ increases.

## 5. Quantitative Results and Discussion.

Thus, it has been demonstrated in the present paper that both the Stockmayer and Flory model overlook important structural elements of the infinite cluster - the mesocycled fragments we managed to visualize as green bonds and quantitatively describe via the density functional approach based on the BMI concept. Such fragments are to be described by a new order parameter $\Theta$, which changes significantly the MAL and weak gel thermodynamics.

To demonstrate, how the MAL is altered due to the MC, we plotted the dependences of the total conversion $\Gamma$ on $\tilde{\rho}_{tot}$ for all the models under discussion for $f=3$[25] and $f=4,5,6,10$ (see Figure 7). They reveal an evident tendency: the more possible structural elements are taken into account, the larger $\Gamma$ is. As seen from Figure 7, the ST model providing the less detailed description of the IC, is getting more and more inadequate as $\tilde{\rho}_{tot}$ and functionality $f$ increase. In turn, the Flory model also underestimates values of $\Gamma$ as compared to both models of mesocycled gel. This underestimation is especially pronounced close to the SGT, the difference being increased as $f$ increases. In particular, the ratio $\Gamma_{mesocycle}^{finite}(\tilde{\rho}_c)/\Gamma_{Flory}(\tilde{\rho}_c)$ is about 1.1 and 1.5 for *f=3* and *f=10*, respectively. The total conversion is seen to depend (in all models but the Flory one) both on the total reduced density $\tilde{\rho}_{tot}$ of the groups *A* and (in the post-gel region) on the monomer functionality *f*. (The analysis for high values of *f* is quite meaningful since the corresponding cluster contribution into the free energy is identical[36] to that for polymer chains each having *f* side FG *A*.)

< Figures 7>

The concentration dependences of some structural characteristics of the gel-fraction, defined in section 4, are presented in Figure 8 for *f=3* and *f=10*. It is seen from Figure 8 that the sol conversion is always lower for the FEA than that calculated at the same reduced group concentration $\tilde{\rho}_{tot}$ for the Flory model disregarding the MC effects.

< Figures 8>

The limiting (in the limit $\tilde{\rho} \to \infty$) values of the index *r* and fraction *w* equals $r_\infty(f)=1-2/f$ and $w_\infty(f) = \lim_{\tilde{\rho}\to\infty} w(\tilde{\rho})$, respectively. When the functionality *f* increases they also increase whereas the limiting internal conversion $\Gamma_i^\infty(f) = \lim_{\tilde{\rho}\to\infty} \Gamma_i(\tilde{\rho})$ decreases. In the limit *f→∞* we find $\Gamma_i^\infty(\infty) = 0.5$ and $w_\infty(\infty) = 1$. Thus, in the completely reacted (Γ=1) weak gel of a high functionality one half of all bonds belongs to mesoscopic cycles and every monomer is included into such cycles via at least two groups *A*. Close to the SGT, $\Gamma_i$ is always higher than *r,* but if *f* is high enough the situation may reverse with increase of $\tilde{\rho}_{tot}$ and Γ (Figure 8).

< Figures 9>

One more view of the gel phase structure is given in Figure 9 where the weight gel fraction $\Phi_g = 1-(\rho_s/\rho)$ and that of the dangling monomers $\Phi_1 = \rho_1/\rho$ are build as functions of $\tilde{\rho}_{tot}$ for *f=3* within all theories. (As discussed in Section 3, in the ST model the IC consists of the dangling monomers only and $\Phi_g=\Phi_1$). Similarly to Figure 7, the more possible structural elements are taken into account, the larger $\Phi_g$ is. One should stress a unique feature of the FEA of the MC model: it predicts that there is no equilibrium gel phase with $\Phi_g<\Phi_{SGT}(f)$. As seen from Figure 9, $\Phi_{SGT}(3)=0.37$. We calculated the function $\Phi_{SGT}(f)$ and found that it smoothly increases with $f$ and approaches a limiting value $\Phi_{SGT}(\infty)=0.48$.

As regards the weight fractions $\Phi_1$ of the dangling (elastically inactive) monomers, it is seen from Figure 9 to pass a maximum and to be the more overestimated by a theory, the less possible structural elements are taken into account by the theory. Indeed, if monomers are included into mesoscopic cycles and/or become junction monomers, they cannot belong to the branches. Notice that the difference between the values of $\Phi_1$ predicted by the Flory and MC theory stays noticeable up to much larger values of the total reduced density $\tilde{\rho}_{tot}$ (and thus total conversion $\Gamma$) than that for $\Phi_g$.

All the data presented in Figures 7-9 clearly evidence that Flory model underestimates the cyclization effects. But the bonding constant $g_0$, appearing in the definition $\tilde{\rho}_{tot} = fg_0\rho$, itself is to be determined from experimental data which makes difficult a direct check of the theoretical predictions (the structural characteristics measured as functions of $\ln\rho$ at different (constant) temperatures should differ only by a shift $\ln\left(fg_0\right)$ along the $\ln\rho$ -axis, though). So, to provide such a comparison for the most aforementioned quantities, we calculated them as functions of the directly measurable total conversion.

< Figures 10>

<Figure 11>

As seen from Figures 10,11, the difference between the Flory and MC model predictions for the quantities defined in both models is less than between both of them and those of the ST model. It stays, though, quite noticeable and disappears only for comparatively large values of $\Gamma$ where both the sol fraction and that of dangling monomers become negligible. The reduced cyclomatic index $r$ is seen to grow here as $\Gamma$ (practically every new bond makes a new cycle since no sol monomers to be attached to the IC are left). Remarkably, even here the MC model predicts a considerable structural change (shift of the cycles-trees equilibrium towards the cycles).

The MC effects, leading to occurrence of the new order parameter $\Theta$ and thus changing the SGT from a geometric event into a genuine phase transition, are even more important for the weak gel thermodynamics. Say, the swallow-tail singularity of the free energy in the vicinity of the SGT, shown in the insets in Figure 6, implies that the SGT line is *always* located within a region where weak gel separates into sol and gel phases.

To make this conclusion even more spectacular, we plotted the pressure, given by eq (2.41), as a function of the concentration for weak gels with concentration dependence of the broken units pressure $P_0(\rho, T) = T\rho$, corresponding to the ideal gas of the broken units, and

$$P_0(\rho, T) = -T\, v^{-1} \ln(1 - v\rho) \tag{5.1}$$

that corresponds to the lattice gas (liquid) of the broken units. For definiteness, we define the lattice cell volume $v$ appearing in (5.1) by the equality $g_0/v = \tilde{\rho}_c$ (a general analysis for an arbitrary values of the ratio $g_0/v$ and the Flory $\chi$-parameter is given elsewhere).

<Figure 12>

As seen from Figure 12, the pressure calculated in the FEA drops stepwise under the IC formation which is due to the stepwise change of the total conversion under this transition caused, in turn, by the stepwise emergence of the order parameter $\Theta$ characterizing MC states. As a result, the function $P(\rho, T)$ splits into two separated branches corresponding to sol and gel phases with no thermodynamically reversible passage between them. We believe that this rather specific feature of the SGT treated with due regard for the MC effects is directly related to the peculiar SGT dynamics observed, e.g. in ref 37.

Thus, we demonstrated in this paper that the effects of mesoscopic cyclization within the gel-fraction are manifold and far from being negligible. They include both considerable quantitative changes of the gel phase structure as compared to the conventional Flory treatment of the tree-like IC and the fundamental change of the SGT itself from a purely geometric phenomenon into a genuine 1st order phase transition. The further improving of our understanding of the IC structure and the SGT nature hardly can be achieved without elaboration a rigorous perturbation procedure to describe the finite cycles within the gel-fraction (which seems to be much harder than that for the sol-fraction[22-25,38]). Until then the only way to check the existing theories of weak gels is to explore and compare all their implications to be observed in real experiments.[39] In such a vein, for both the Flory and MC models we compared a global weak gel phase behavior (the classification of their phase diagrams)[36,41] and studied the coil-globule transition in the associating solvent[42]. The scattering singularities due to the mesoscopic cyclization effects are discussed in refs 23 and 43 within the IEA and FEA, respectively.

The authors acknowledge the financial support by INTAS (grant INTAS 99-01852), DFG (SFB 481) and the Russian governmental program "Universities of Russia (Basic Researches)". I.E. is grateful to W. Stockmayer and S. F. Edwards for encouraging comments and to S. Panyukov, I. Nyrkova and A. Semenov for fruitful discussions.

**Appendix. Free energy and entropy of a dimeric system.** Let $N$ dimers $AB$ be confined in the volume $V$ and exposed to the external fields $\phi_A(\mathbf{r})$ and $\phi_B(\mathbf{r})$ applied to the monomers $A$ and $B$ the dimers are formed of. It follows from definitions (2.3)-(2.5) that the function of distribution in the configuration space and the free energy of the system are

$$f(\mathbf{X}) = \exp\left(-\sum_{i=1}^{i=N}\left(\phi_A(\mathbf{r}_{2i-1}) + \phi_B(\mathbf{r}_{2i})\right)\Big/T\right)\prod_{i=1}^{i=N} g(\mathbf{r}_{2i-1} - \mathbf{r}_{2i}) \qquad (A1)$$

$$F_{AB}\left(N, V, T, \{\phi_A(\mathbf{r})\}, \{\phi_B(\mathbf{r})\}\right) = -NT\ln\left(e\int\psi_A\,\hat{g}\psi_B\,dV\Big/N\right) \qquad (A2)$$

with $\psi_i(\mathbf{r}) = \exp(-\phi_i(\mathbf{r})/T)$ and $\left(\hat{g}\psi_i\right)(\mathbf{r}) = \int d\mathbf{r}'\,g(\mathbf{r} - \mathbf{r}')\psi_i(\mathbf{r}')$. On the other hand

$$F_{AB}\big(N,V,T,\{\phi_A(\mathbf{r})\},\{\phi_B(\mathbf{r})\}\big)=E\big(\{\phi_i(\mathbf{r})\},\{\overline{\rho}_i(\mathbf{r})\}\big)-TS_{AB}\big(\{\overline{\rho}_i(\mathbf{r})\}\big) \qquad (A3)$$

Here $E\big(\{\phi_i(\mathbf{r})\},\{\rho_i(\mathbf{r})\}\big)=\displaystyle\sum_{i=A,B}\int dV\,\rho_i(\mathbf{r})\,\phi_i(\mathbf{r})$ is the energy of the system with certain non-uniform density profiles $\{\rho_i(\mathbf{r})\}$ of the monomers subjected to the external fields $\{\phi_i(\mathbf{r})\}$, $\overline{\rho}_i(\mathbf{r})$ are the density profiles equilibrium in the applied external fields $\phi_i(\mathbf{r})$ and $S_{AB}\big(\{\overline{\rho}_i(\mathbf{r})\}\big)$ is the entropy of the system with such density profiles, $\overline{\rho}$ and $\phi$ being interrelated as follows:

$$\phi_i(\mathbf{r})=T\,\delta S_{AB}\big(\{\overline{\rho}_i(\mathbf{r})\}\big)/\delta\overline{\rho}_i(\mathbf{r}),\quad \overline{\rho}_i(\mathbf{r})=-\delta F_{AB}\big(\{\phi_i(\mathbf{r})\}\big)/\delta\phi_i(\mathbf{r}) \qquad (A4)$$

It follows from (A2)-(A4) that

$$\overline{\rho}_A(\mathbf{r})=-\frac{\delta F_{AB}}{\delta\phi_A(\mathbf{r})}=N\,\frac{\psi_A(\mathbf{r})\hat{g}\psi_B}{\int\psi_A\hat{g}\psi_B\,dV},\quad \overline{\rho}_B(\mathbf{r})=-\frac{\delta F_{AB}}{\delta\phi_B(\mathbf{r})}=N\,\frac{\psi_B(\mathbf{r})\hat{g}\psi_A}{\int\psi_A\hat{g}\psi_B\,dV} \qquad (A5)$$

Substituting (A5) into (A3) and the definition of $E\big(\{\phi_i(\mathbf{r})\},\{\rho_i(\mathbf{r})\}\big)$, we get finally the desired expressions for the entropy of the system of dimers and that of the directed bond formation:

$$S_{AB}\big(\{\rho_A(\mathbf{r})\},\{\rho_B(\mathbf{r})\}\big)=\frac{1}{2}\left(\int dV\,\rho_A(\mathbf{r})\ln\frac{e\hat{g}\psi_B}{\rho_A\psi_A}+\int dV\,\rho_B(\mathbf{r})\ln\frac{e\hat{g}\psi_A}{\rho_B\psi_B}\right) \qquad (A7)$$

$$\begin{aligned}S_{bond}^{AB}\big(\{\rho_A(\mathbf{r})\},\{\rho_B(\mathbf{r})\}\big)&=S_{AB}-\int dV\,\rho_A(\mathbf{r})\ln\big(e/\rho_A(\mathbf{r})\big)-\int dV\,\rho_B(\mathbf{r})\ln\big(e/\rho_B(\mathbf{r})\big)\\ &=\frac{1}{2}\int dV\left(\rho_A(\mathbf{r})\ln\frac{\rho_A\,\hat{g}\psi_B}{e\psi_A}+\rho_B(\mathbf{r})\ln\frac{\rho_B\hat{g}\psi_A}{e\psi_B}\right)\end{aligned} \qquad (A8)$$

with functions $\rho_i(\mathbf{r})$, $\psi_i(\mathbf{r})$ from (A6). For the uniform distribution ($\rho_A(\mathbf{r})=\rho_B(\mathbf{r})=const$) the entropy of the directed bond formation is simply

$$S_{bond}^{AB}\big(\rho_A,\rho_B\big)=S_{bond}^{AB}\big(\rho_A,\rho_A\big)=V\rho_A\ln\big(g_0\rho_A/e\big) \qquad (A9)$$

value of the negative (attraction) energy they acquire if locating in the neighboring sites of the lattice. So, these two models could belong to different universality classes.

# Figure Captions.

**Figure 1**. The typical blocks serving as bricks under formation of clusters of thermoreversibly bonded $A_3$ monomers. a) - single monomer ("bare" vertex), b,c) - the simplest cycled blocks of 1 and 2 independent cycles, respectively, d) - a more complicated cycled block consisting of 6 independent cycles. The side "shoots" correspond to both the unreacted groups $A$ and to those that form "external" bonds connecting the blocks into a larger cluster.

**Figure 2.** A typical finite fragment of the gel phase with the proper coloration of FG and monomers induced by the monomer identity breaking as consistent with the Flory model of gel formed of $A_3$ monomers.

**Figure 3.** The full list of the monomer coloration types for $f=3$: a) $S_3$ (sol), b) $D^+D_2$ (dangling), c) $DI_2$ (backbone), d) $I_3$ (tree junction) e) $DC_2$ (mesocycle backbone), f) $IC_2$ (mesocycle-tree junction) and g) $C_3$ (mesocycle junction) monomers. The Stockmayer, Flory, IEA and FEA of the mesocycled gel model take into account a)-c), a)-d), a)-f) and a)-g) types, respectively.

**Figure 4.** Visualization of the BMI concept for the MC models. a) A typical finite fragment of the gel phase with the proper coloration of FG and monomers induced by the monomer identity breaking. A considerable number of small and simple cycles seen here are due to the fact that the condition $\varepsilon = \left( f\rho a^3 \right)^{-1} \ll 1$ can not be satisfied in the 2D space. b), c) - the different colorations of a smaller IC fragment (delineated by the dashed rectangle in Figure 4a) as consistent with the Flory and MC models, respectively.

**Figure 5.** The MC increment of the free energy $\varphi_{meso}$ as a function of the fraction of bonds involved into mesoscopic cycles at different values of the reduced group density $\tilde{\rho}_{tot}$ for the IEA ($a$) and FEA ($b,c$). In $a$) and $b$) $\tilde{\rho}_{tot} = \tilde{\rho}_c + (i-3)*0.05$, where $i$ are the numbers labeling the curves, in $c$) $\tilde{\rho}_{tot} = \tilde{\rho}_{SGT} + (i-4)(\tilde{\rho}_{SGT} - \tilde{\rho}_{min})/2$. The behavior of the curves plotted in Figure 5b for low values of the internal conversion $\Gamma_i$ is shown within the inset.

**Figure 6.** The dependences of the cluster contributions into equilibrium structural free energy of the total reduced group density $\tilde{\rho}_{tot}$ for Stockmayer and Flory models (the dashed and bold lines, respectively) and those for mesoscopic cyclization models (the dotted and solid lines, respectively). To visualize their distinction near the sol-gel threshold where these contributions are rather close to each other, the increments $\varphi = \left(F_{mesocycle}(\tilde{\rho}_{tot}) - F_{Flory}(\tilde{\rho}_{tot})\right)/T$ and $\left(F_{Stockmayer}(\tilde{\rho}_{tot}) - F_{Flory}(\tilde{\rho}_{tot})\right)/T$ are plotted in the insets. *a) f=3, b) f=10.*

**Figure 7.** The total conversion $\Gamma$ as a function of $\tilde{\rho}_{tot}$ for the Flory and Stockmayer models (bold and dotted lines, respectively) and the MC model in the IEA and FEA (dashed and solid lines, respectively). The curves are plotted for the values of the monomer functionality *f=4,5,6,10* and labeled *1,2,3,4,* respectively. All the curves corresponding to the same value of *f* start from the same point, *which* enables to label only one (Stockmayer) curve from every set. For *f=6* a vicinity of the SGT is shown in the inset.

**Figure 8a.** The total ($\Gamma$), sol ($\Gamma_s$), external ($\Gamma_e$) and internal ($\Gamma_i$) conversions (the curves *1,2,3,4* respectively), reduced cyclomatic index *r* (the curve *5*) and fraction *w* of monomers included into mesoscopic cycles (the curve *6*) as functions of $\tilde{\rho}_{tot}$ calculated for *f=3* in FEA (solid lines). The quantities $\Gamma$, $\Gamma_s$ and *r* are calculated also for the Flory model (dashed lines) and labeled similarly.

**Figure 8b.** The same plots for *f=10.* A vicinity of the SGT is shown in the inset.

**Figure 9.** The weight gel fraction $\Phi_g$ (solid lines) and that of dangling monomers $\Phi_1$ (dashed lines) as functions of $\tilde{\rho}_{tot}$ for *f=3* calculated for Stockmayer (*1*), Flory (*2*) and mesoscopic cyclization models in the IEA (*3*) and FEA (*4*). The asterisks indicate the location of the first order SGT on the FEA curves.

**Figure 10.** The structural characteristics of the gel phase as functions of the total conversion $\Gamma$ for *f=3.* a) The straight line $\Gamma=\Gamma$ (*1*) as a baseline and $\Gamma_s$, $\Gamma_e$, *r* and *w* (the curves *2,3,5,6,*

respectively) calculated in FEA (solid lines). The sol conversion $\Gamma_s$ and $r$ are calculated also for the Flory model (dashed lines) and labeled similarly. The dotted line *2* corresponds to the Stockmayer model result for $\Gamma_s$. b) The fraction of the bonds involved into mesoscopic cycles (internal conversion) $\Gamma_i$ (the curve *4*), weight gel fraction $\Phi_g$ (the curves *7*) and that of dangling monomers $\Phi_1$ (the curves *8*). The results calculated in FEA and for Stockmayer and Flory models are plotted as the solid, dotted and dashed lines, respectively.

**Figure 11.** The same plots as in Figure 10 for *f=10*.

**Figure 12.** The weak gel pressure as a function of its volume fraction $\phi$ for *f*=3. a) The cluster contribution to pressure for Flory and Stockmayer models (bold and dotted lines, respectively) and the MC model in the IEA and FEA (dashed and solid lines, respectively). b) A typical isotherm of the lattice weak gel with a properly chosen temperature. The asterisks indicate the limits of stability for the sol (tree-like) and gel (mesocycled) phases at a fixed value of $\phi$.

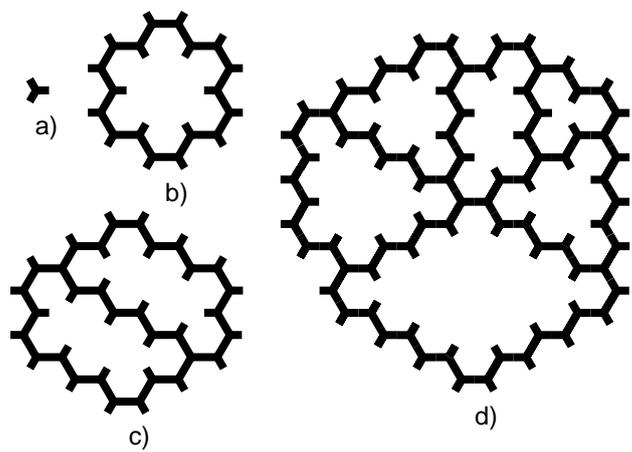

a)

b)

c)

d)

Figure 1.

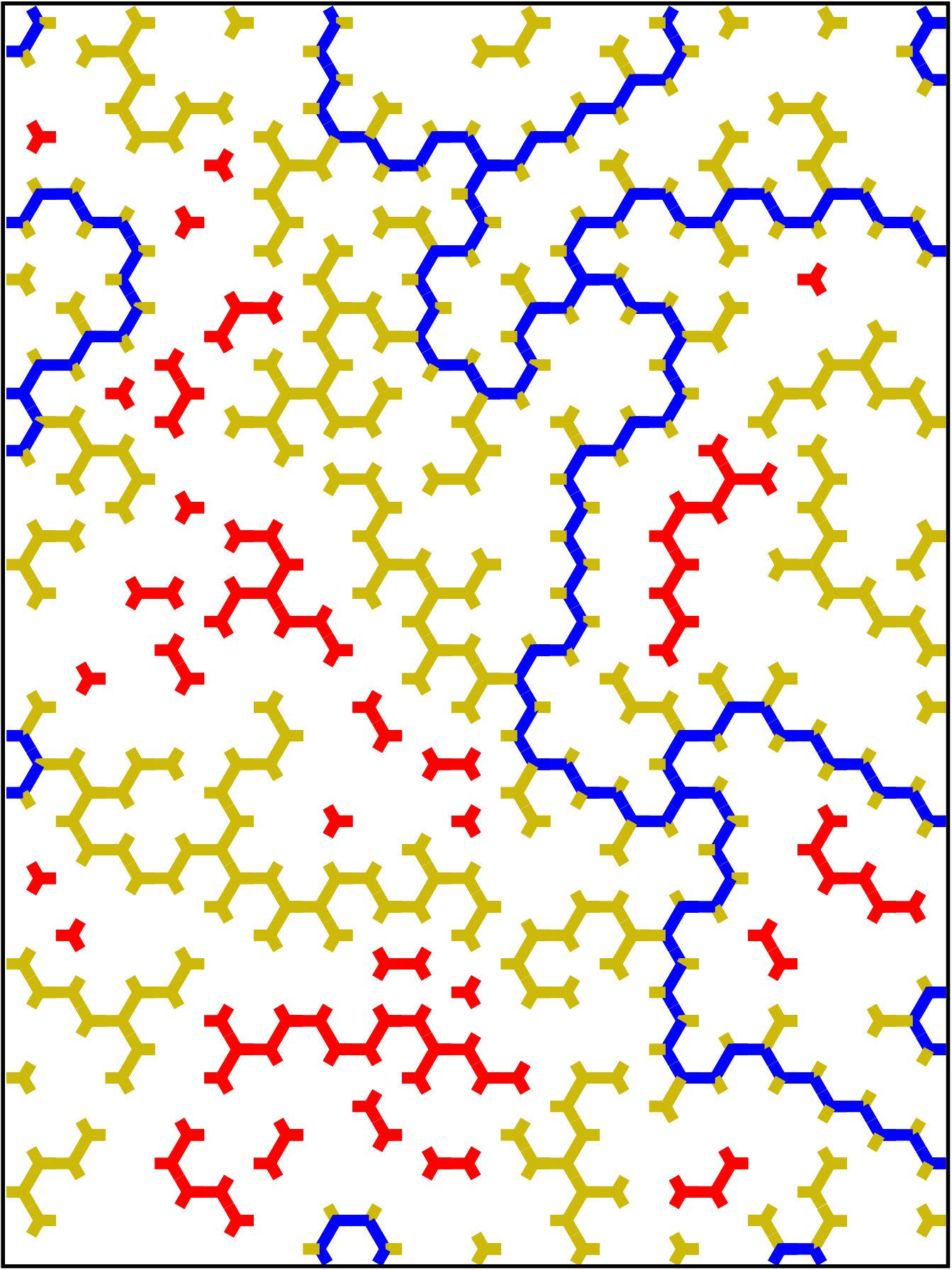

Figure 2.

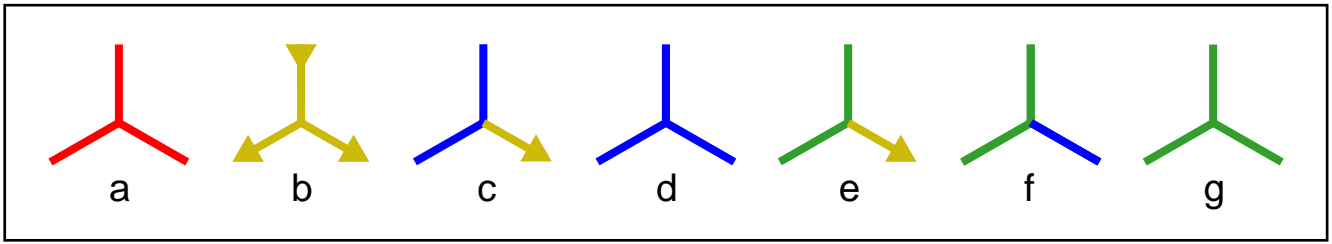

Figure 3.

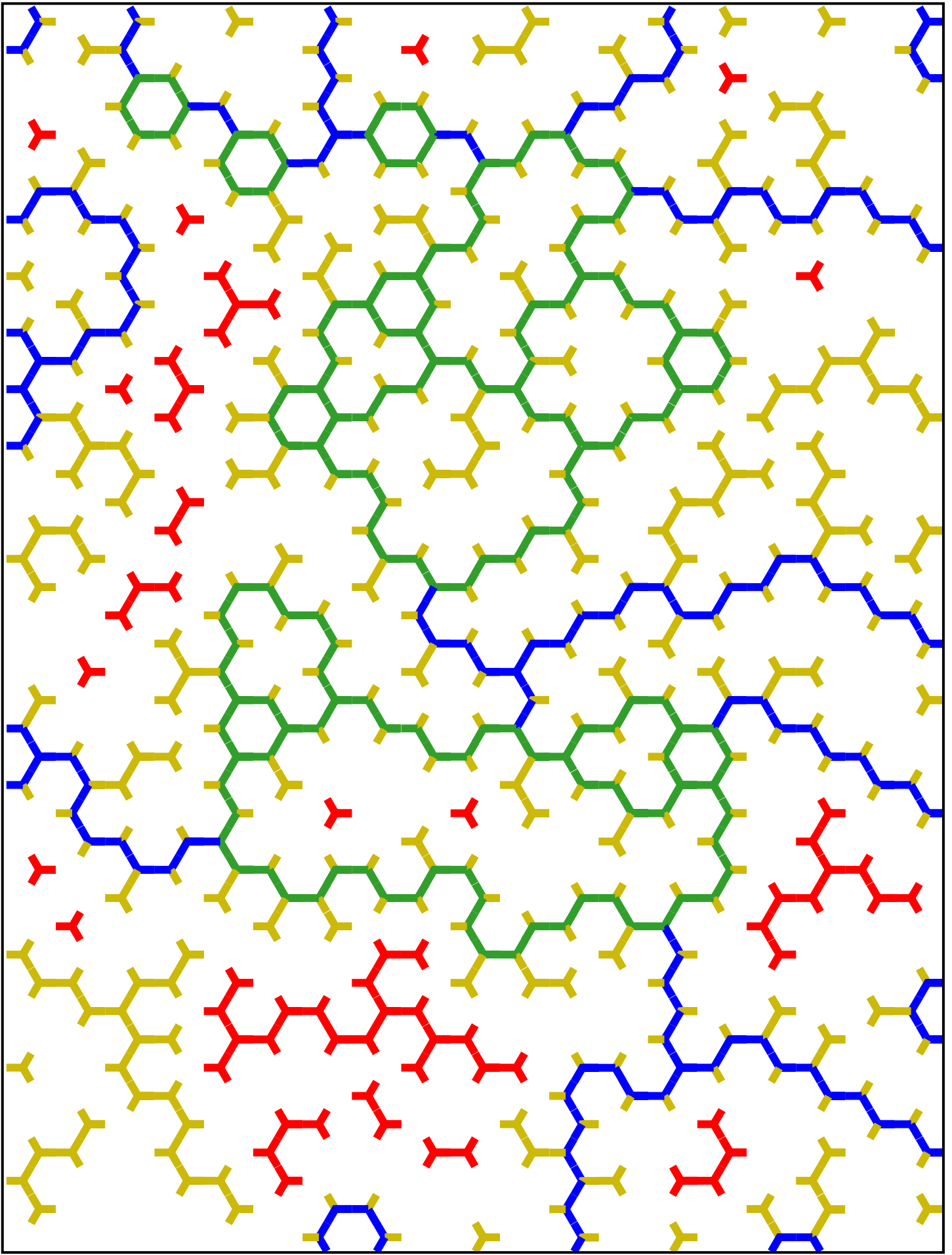

Figure 4a.

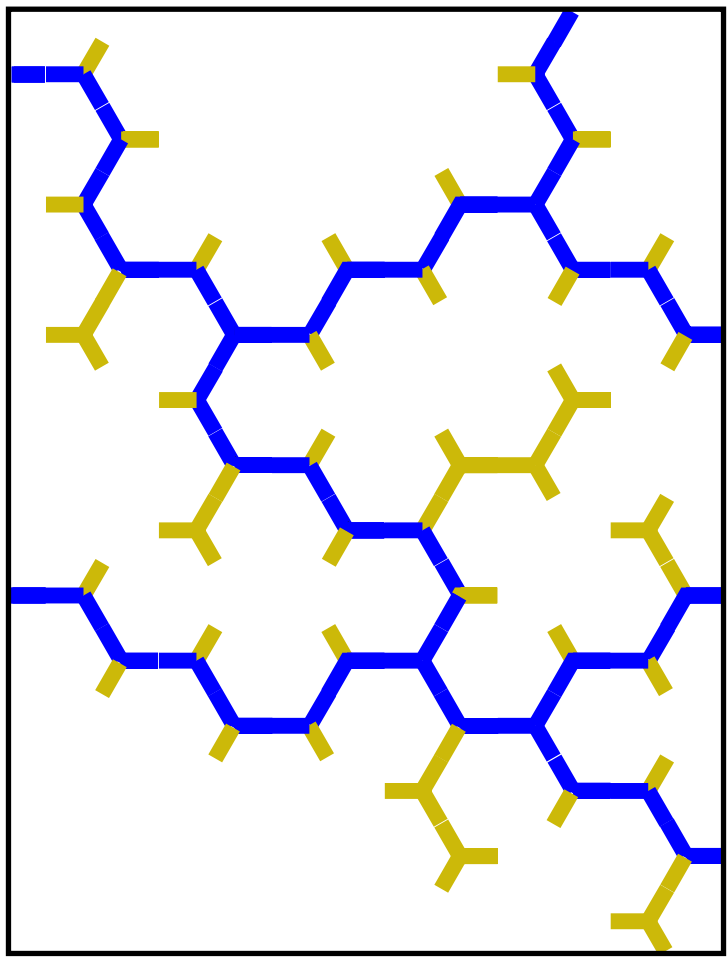

Figure 4b.

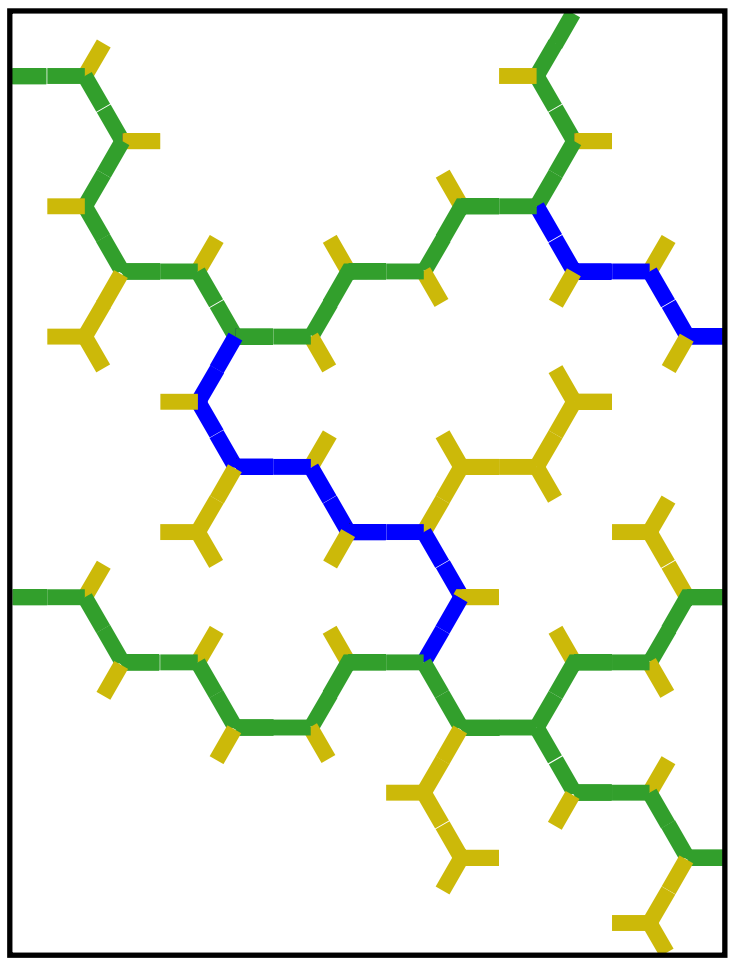

Figure 4c.

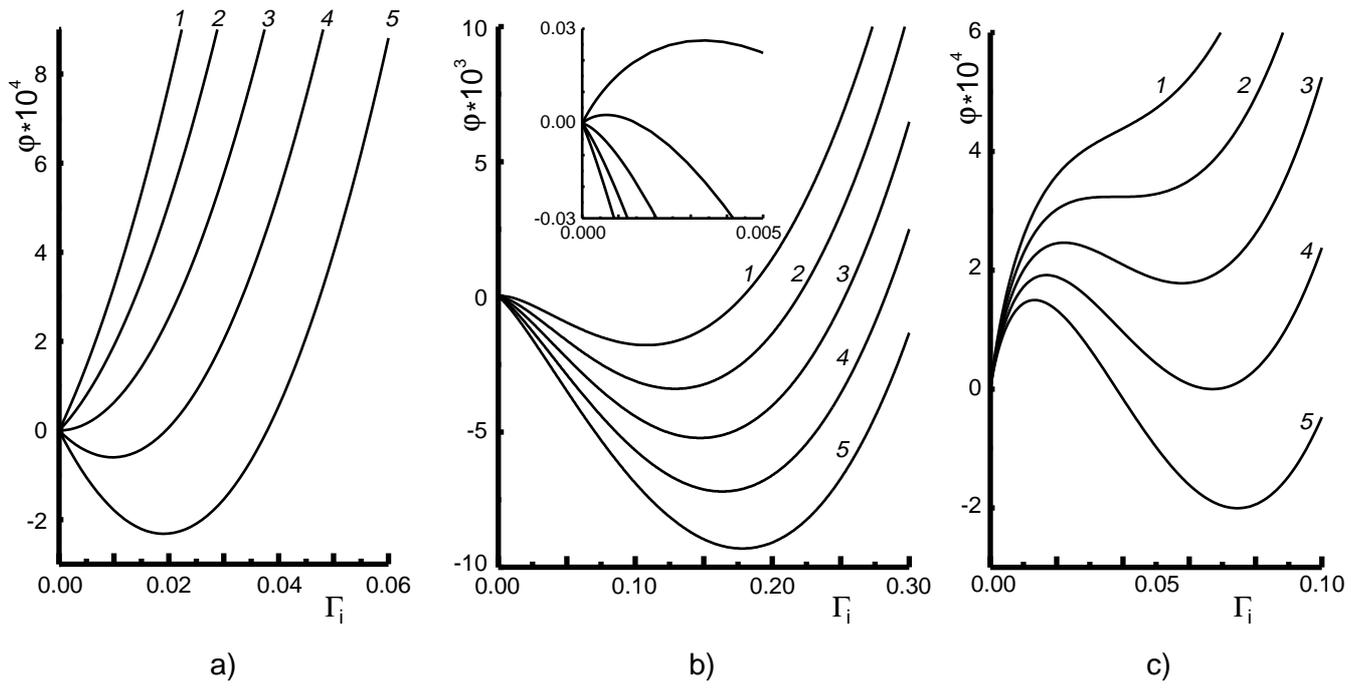

Figure 5.

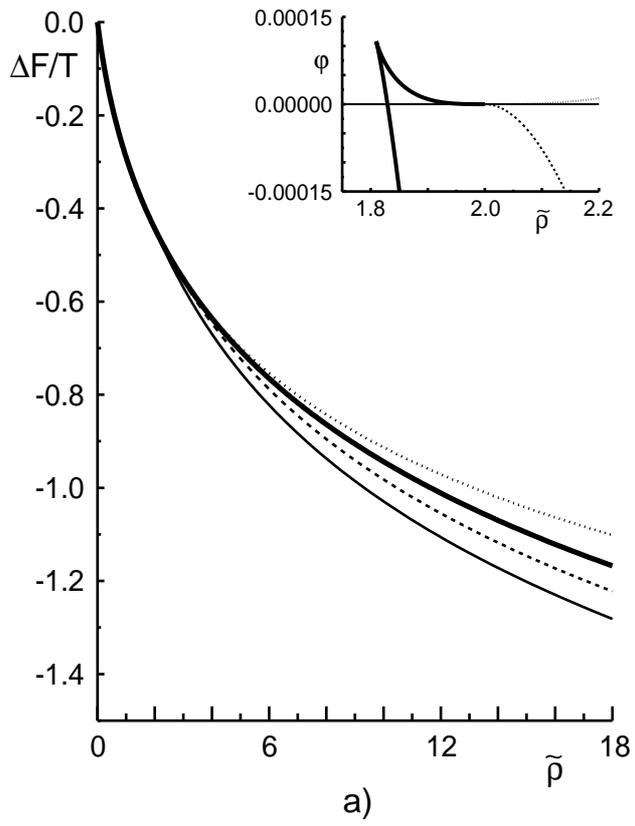

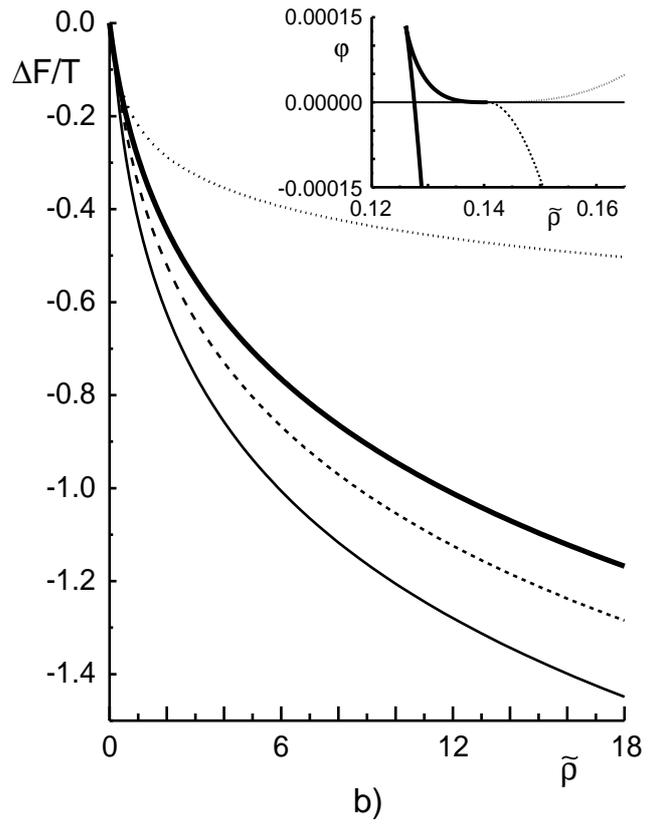

a)

b)

Figure 6.

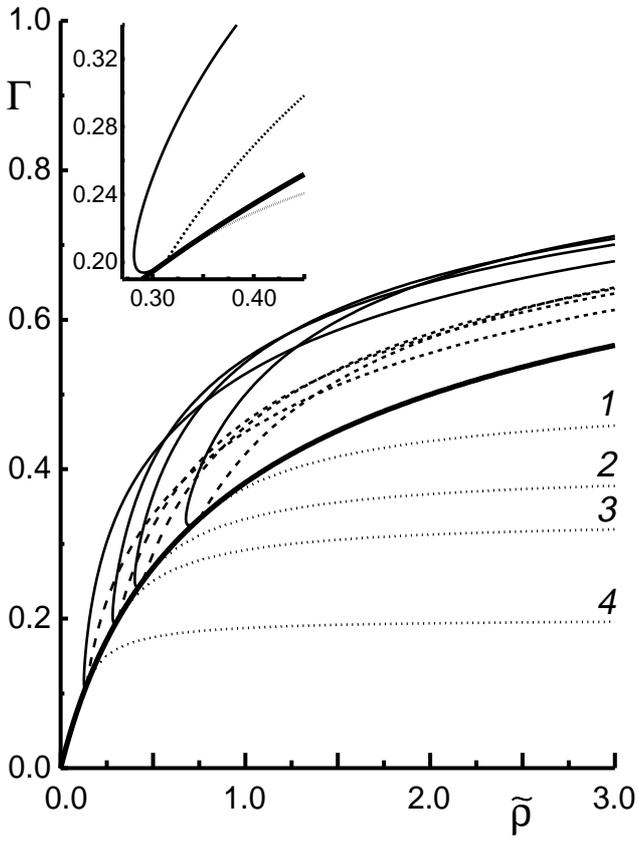

Figure 7.

Figure 8.

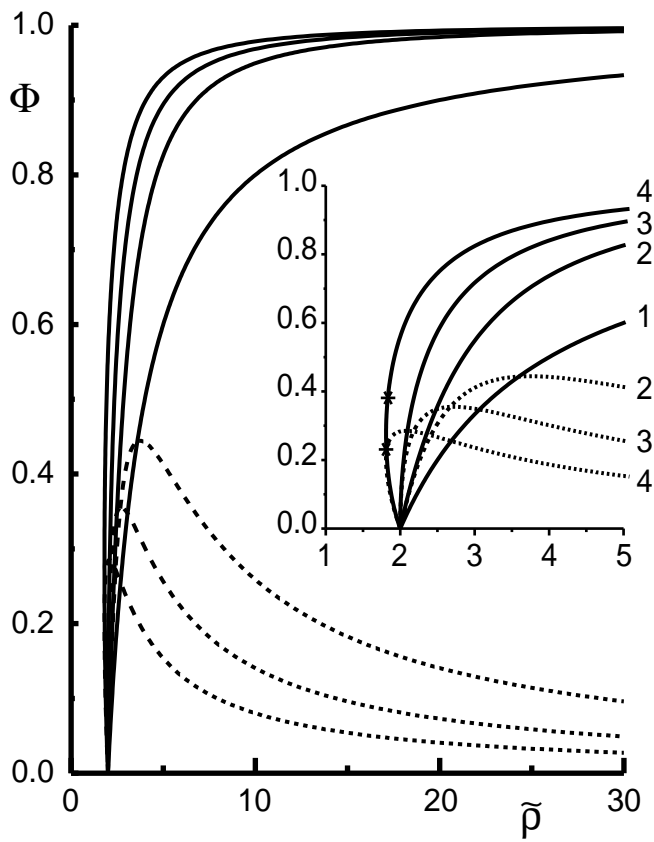

Figure 9.

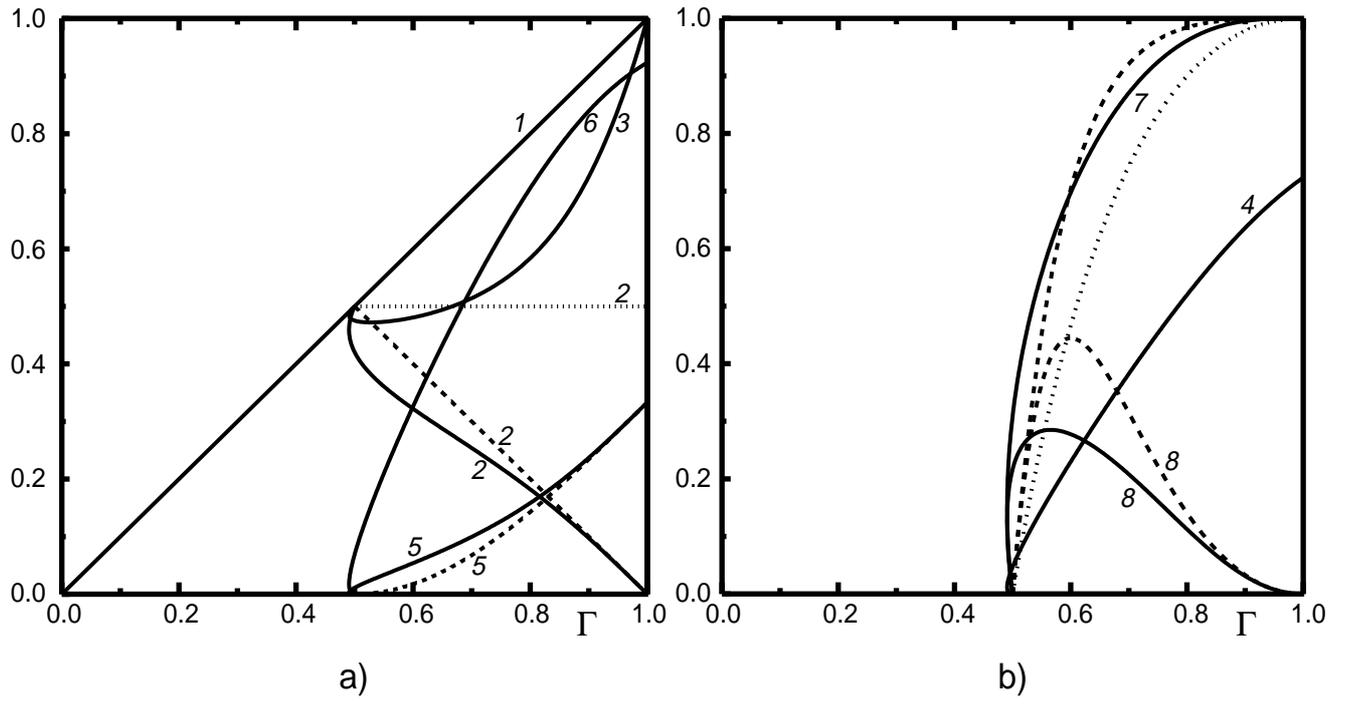

Figure 10.

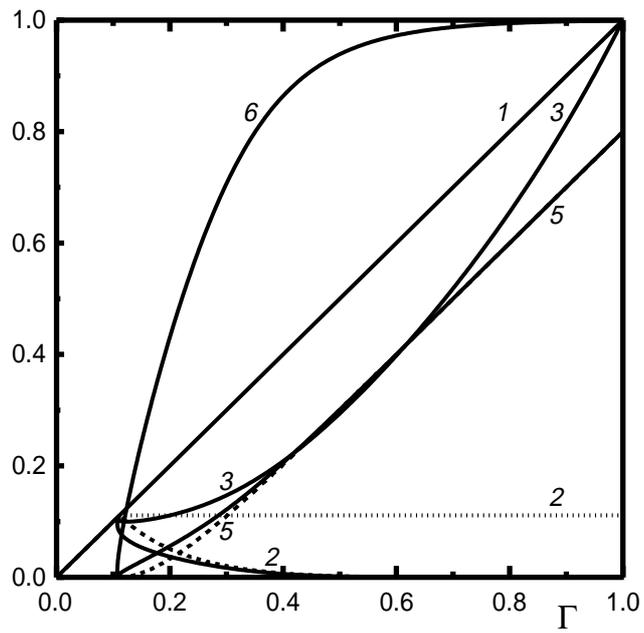 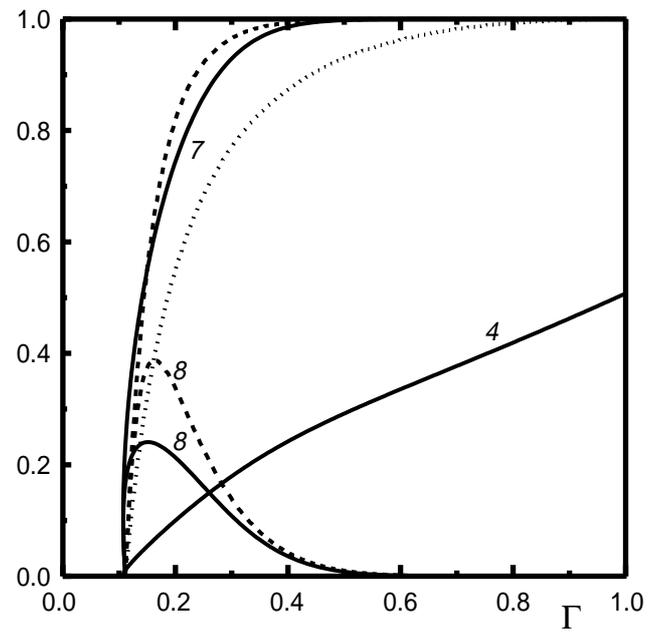

a)                                      b)

Figure 11.

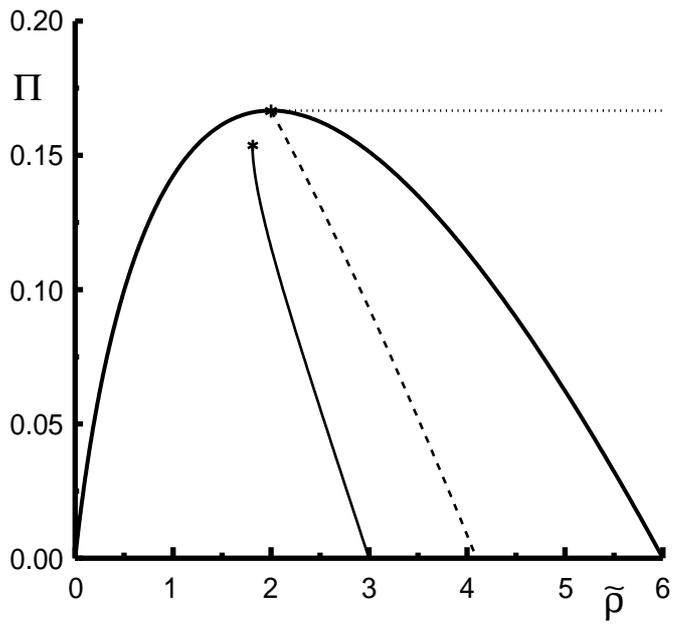
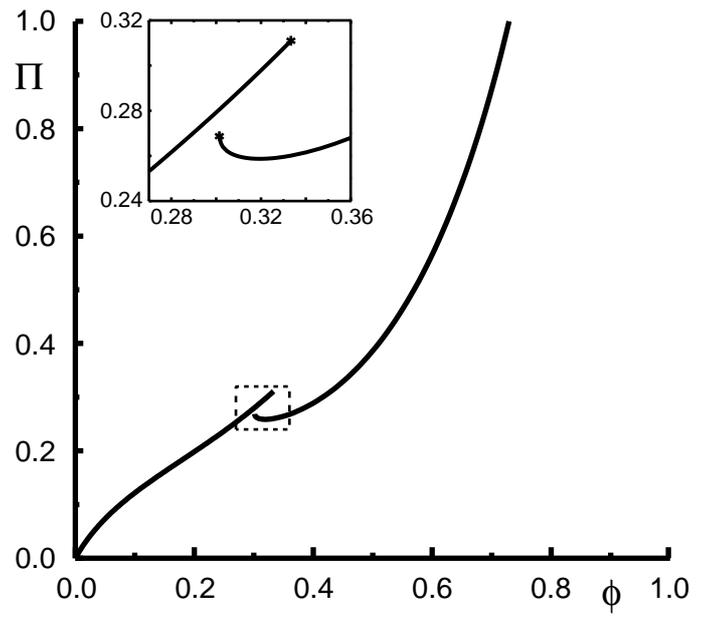

a)

b)

Figure 12.